\documentclass[11pt]{article}

\usepackage{amssymb,amsmath}
\usepackage[dvips]{graphicx}
\usepackage{subfigure}
\usepackage{latexsym}
\usepackage[dvips]{graphicx}

\def\abstract#1{\vskip 7mm 
        \begin{center}{\large Abstract}\par \smallskip
                \begin{minipage}[c]{15.5cm}
                       #1
                \end{minipage}
        \end{center}
}
\def\title#1{\begin{center}{\Large\bf #1}\end{center}}
\def\author#1{\vskip 5mm \begin{center}{#1}\end{center}}
\def\address#1{\begin{center}{\it #1}\end{center}}

\numberwithin{equation}{section}

\newcommand{\eqb}{\begin{equation}}
\newcommand{\eqe}{\end{equation}}
\newcommand{\eqbnon}{\begin{equation*}}
\newcommand{\eqenon}{\end{equation*}}

\newcommand{\eqab}{\begin{eqnarray}}
\newcommand{\eqae}{\end{eqnarray}}
\newcommand{\eqabnon}{\begin{eqnarray*}}
\newcommand{\eqaenon}{\end{eqnarray*}}

\newcommand{\seqb}{\begin{subequations}}
\newcommand{\seqe}{\end{subequations}}

\newcommand{\eref}[1]{\eqref{#1}} % should be excluded in using ``iopart''

\newcommand{\defeq}{:=}
\newcommand{\defeqr}{=:}

\newcommand{\pd}[2]{\dfrac{\partial #1}{\partial #2}}

\newcommand{\ad}[2]{{\mathcal Ad}~: #1 \rightsquigarrow #2}
\newcommand{\revad}[2]{{\mathcal Ad_{rev}}~: #1 \leftrightsquigarrow #2}

\newcommand{\bra}[1]{\left< #1 \right|}
\newcommand{\ket}[1]{\left| #1 \right>}
\newcommand{\bracket}[1]{\left< #1 | #1 \right>}

\newtheorem{define}{Definition}
\newtheorem{fact}{Fact}
\newtheorem{thm}{Theorem}
\newtheorem{prop}{Proposition}
\newtheorem{lemma}{Lemma}
\newtheorem{coro}{Corollary}

\begin{document}

\title{Universal Property of Quantum Gravity\\ implied by\\ Uniqueness Theorem of Bekenstein-Hawking Entropy
\footnote{
Invited as a feature paper, and accepted as a refereed paper, for the special issue {\em Black Hole Thermodynamics} in the journal {\em Entropy}, edited by J.Bekenstein.
}}

\author{Hiromi Saida~\footnote{Email : saida@daido-it.ac.jp}}
\address{Department of Physics, Daido University, 10-3 Takiharu Minami-ku, Nagoya 457-8530, Japan}
%\ead{saida@daido-it.ac.jp}

\abstract{
This paper consists of three steps.
In the first, we prove that the Bekenstein-Hawking entropy is the unique expression of black hole entropy.
Our proof is constructed in the framework of thermodynamics without any statistical discussion.
In the second, intrinsic properties of quantum mechanics are shown, which justify the Boltzmann formula to yield a unique entropy in statistical mechanics. 
These properties clarify three conditions, one of which is necessary and others are sufficient for the validity of Boltzmann formula. 
In the third, by combining the above results, we find a reasonable suggestion from the sufficient conditions that the potential of gravitational interaction among microstates of underlying quantum gravity may not diverge to negative infinity (such as Newtonian gravity) but is bounded below at a finite length scale. 
In addition to that, from the necessary condition, the interaction has to be repulsive within the finite length scale. The length scale should be Planck size. 
Thus, quantum gravity may become repulsive at Planck length.
Also, a relation of these suggestions with action integral of gravity at semi-classical level is given. 
These suggestions about quantum gravity are universal in the sense that they are independent of any existing model of quantum gravity.
% such as superstring theory, loop quantum gravity, (causal) dynamical triangulation and so on.
}

%\twocolumn
% \tableofcontents
%\onecolumn

%%%%%%%%%%%%%%%%%%%%%%%%%%%%%%%%%%%%%%%%%%%%%%%%%%%%%%%%%%%%%%%%%%%%%%%%%%%%%%%%%%%%%%%%%%%%%%%%%%%%
\section{Introduction}
\label{sec:intro}

Gravity is the only fundamental interaction which is not quantized at present.
By combining classical physics (general relativity) of black holes and quantum field theory in black hole spacetime, it is theoretically very reasonable to regard the stationary black hole as a thermal equilibrium state of gravity whose temperature is determined by the thermal spectrum of Hawking radiation~\cite{ref:bardeen+2.1973,ref:bekenstein.1973,ref:bekenstein.1974,ref:braden+3.1990,ref:brown+2.1991,ref:davies.1977,ref:hawking.1971,ref:hawking.1975,ref:israel.1986,ref:york.1986}.
This theoretical evidence gives us the notion of {\em black hole thermodynamics} (see App.\ref{app:free}).
When a thermal system includes {\em single} black hole, the entropy of black hole ({\em Bekenstein-Hawking entropy}) is given by the entropy-area law, which claims that the equilibrium entropy of event horizon is equal to one-quarter of its spatial area in Planck units~\cite{ref:bekenstein.1973,ref:bekenstein.1974,ref:flanagan+2.2000,ref:frolov+1.1993,ref:saida.2006,ref:unruh+1.1982,ref:unruh+1.1983}.
(Note that, for a multi-event horizon system, the Bekenstein-Hawking entropy is not necessarily given by the entropy-area law~\cite{ref:saida.2009b}.
Thus, in this paper, the term ``Bekenstein-Hawking entropy'' denotes simply the entropy of black hole, and it is distinguished from the term ``entropy-area law''.)
However, at present, the black hole thermodynamics is nothing more than a conjecture in the sense that the microstate responsible for Bekenstein-Hawking entropy is unknown.

It is reasonable to consider that the microstates composing black hole are microstates of underlying quantum gravity, since the Hawking radiation~\cite{ref:birrell+1.1982,ref:hawking.1975}, which is the key theoretical evidence of black hole thermodynamics, is a significant prediction of quantum field theory in curved spacetime.
Then, we expect that some quantum property of gravity is extracted by studying the microscopic origin of Bekenstein-Hawking entropy.
Many of existing researches on Bekenstein-Hawking entropy seem to consider mainly a relation between spacetime-geometric aspects and microscopic meanings of the entropy (e.g. see~\cite{ref:bousso.2002} for a notable suggestion given by those geometrical considerations).
On the other hand, it seems that thermodynamic and statistical mechanical foundations of Bekenstein-Hawking entropy remain to be examined rigorously:

In order to consider thermodynamic foundation of Bekenstein-Hawking entropy, we note that the basic principles of ordinary thermodynamics for laboratory systems are not only the four laws of thermodynamics but also, for example, the intensivity and extensivity of state variables, the additivity of extensive variables, the existence of adiabatic process, and so on.
(In axiomatic formulation of ordinary thermodynamics for laboratory systems, there are some other basic principles~\cite{ref:lieb+1.1999,ref:tasaki.2000}.)
Those basic principles of ordinary thermodynamics result in, for example, the {\em uniqueness of entropy}, thermal stability of thermodynamic system and so on.
(Here, the ``uniqueness'' means that any state variable, $K$, satisfying extensivity, additivity and so-called entropy principle whose detail are given in Sec.\ref{sec:td}, is necessarily related to the entropy, $S$, as, $K = \alpha\,S + \eta$, where $\alpha$ and $\eta$ are suitable constants.) 
However, because some basic principles of ordinary thermodynamics are not retained in black hole thermodynamics as shown in Sec.\ref{sec:td}, the uniqueness of Bekenstein-Hawking entropy, for example, is not necessarily manifest.

Next, in order to consider statistical mechanical foundation of Bekenstein-Hawking entropy, we note that the Boltzmann formula in ordinary quantum statistical mechanics is justified by some properties of quantum mechanics.
An example of the property is the existence of unique thermodynamic limit of logarithmic density of number of states, $\lim_{t.l.}V^{-1}\ln\Omega$, where $V$ is volume of system, $\Omega$ is number of states and $\lim_{t.l.}$ means thermodynamic limit (i.e. $V\to\infty$ with fixing the energy density and particle number density at finite values)~\cite{ref:dobrushin.1964,ref:ruelle.1999,ref:tasaki.2008}.
This property of quantum mechanics ensures the existence of {\em unique thermodynamic limit of entropy density}, $\lim_{t.l.}S/V$, defined by Boltzmann formula, $S \defeq k_B\ln\Omega$.
If the Bekenstein-Hawking entropy is given by the Boltzmann formula, this property of quantum mechanics may be related with some property of underlying quantum gravity.

This paper examines thermodynamic and statistical mechanical foundations of Bekenstein-Hawking entropy, and gives a reasonable suggestion about the gravitational interaction among microstates of underlying quantum gravity.
In Sec.\ref{sec:td}, the thermodynamic foundation is examined, where some basic principles of ordinary thermodynamics are modified in black hole thermodynamics, and then the uniqueness of Bekenstein-Hawking entropy is proven.
Its proof is constructed in the framework of thermodynamics without any statistical discussion.

Sec.\ref{sec:bf} concerns the intrinsic properties of quantum mechanics which justify the Boltzmann formula.
In that section, we clarify the sufficient conditions (conditions~A and~B in theorem~\ref{thm:rt}) satisfied by the interaction potential among constituent quantum particles, so as to ensure the existence of unique thermodynamic limit, $\lim_{t.l.}V^{-1}\,\ln\Omega$, which justifies the Boltzmann formula as mentioned in previous paragraph.
One of the sufficient conditions is that the interaction potential has a negative lower bound at a finite length scale, $R_{\rm bound}$. 
(Note that, at least for laboratory systems, there seems to be no example which violates this sufficient condition but retains the Boltzmann formula.) 
Also in Sec.\ref{sec:bf}, we show the necessary condition for the existence of thermal equilibrium states of quantum system, for the case that the interaction among many particles is a sum of two-particle interactions and multi($\ge 3$)-particle interactions do not exist. 
The necessary condition is that the two-particle interaction potential has to be large positive in a suitable region. 
(Note that the validity of Boltzmann formula and the existence of thermal equilibrium states are separately considered.) 
Then it is found in that section that, for the quantum system satisfying the necessary condition for the existence of thermal equilibrium states and the sufficient conditions for the validity of Boltzmann formula, the two-particle interaction potential should be repulsive within the length scale $R_{\rm bound}$.

Finally, Sec.\ref{sec:conc} is for the conclusion:
Let us adopt two suppositions; (i) the stationary black hole is a thermal equilibrium state of microstates of underlying quantum gravity, and (ii) statistical mechanics is applicable to the black hole.
Under these suppositions, the {\em uniqueness} of Bekenstein-Hawking entropy shown in Sec.\ref{sec:td} implies that the Boltzmann formula, which yields the {\em unique} entropy, is valid even for the underlying quantum gravity. 
Then, since no counter-example to the sufficient conditions seems to be found at least in laboratory systems, it seems to be empirically reasonable that the sufficient conditions for the validity of Boltzmann formula hold also in quantum gravity. 
In this case, the interaction potential among microstates of underlying quantum gravity is bounded below at a finite length scale $R_{\rm bound}$, unlike the Newtonian gravity. 
On the other hand, the existence of thermal equilibrium state (black hole) implies that the underlying quantum gravity satisfies the necessary condition for the existence of thermal equilibrium states of laboratory quantum system, which means that the potential of two-body gravitational interaction becomes large positive in a suitable region. 
Thus, when the underlying quantum gravity satisfies the necessary condition for the existence of thermal equilibrium states and the sufficient conditions for the validity of Boltzmann formula, the two-body interaction should be repulsive within the length scale $R_{\rm bound}$. 
This $R_{\rm bound}$ may be the Planck length, at which the quantum gravitational effect appears significantly. 
That is, the quantum gravity may become repulsive at Planck length. 
Moreover, in Sec.\ref{sec:conc}, a relation of these suggestions with action integral of gravity at semi-classical level is also given. 
Those suggestions about quantum gravity are universal in the sense that they are independent of any existing model of quantum gravity (e.g. superstring theory, loop quantum gravity, (causal) dynamical triangulation, and so on), since discussions in this paper do not use any existing model of quantum gravity.

Minimal reviews of important topics are given in some sections; axiomatic thermodynamics in Subsec.\ref{sec:td.axiom}, black hole thermodynamics in Subsec.\ref{sec:td.bh} and rigorous foundation of quantum statistical mechanics in Sec.\ref{sec:bf}.
A reader who knows the topic can skip the corresponding review section.

%%%%%%%%%%%%%%%%%%%%%%%%%%%%%%%%%%%%%%%%%%%%%%%%%%%%%%%%%%%%%%%%%%%%%%%%%%%%%%%%%%%%%%%%%%%%%%%%%%%%
\section{Uniqueness of Bekenstein-Hawking Entropy}
\label{sec:td}

%%%%%%%%%%%%%%%%%%%%%%%%%%%%%%%%%%%%%%%%%%%%%%%%%%
\subsection{Ordinary Thermodynamics in Axiomatic Formulation}
\label{sec:td.axiom}

The best preparation for the aim of this paper may be a review of the whole of axiomatic thermodynamics~\cite{ref:lieb+1.1999,ref:tasaki.2000}.
However in this subsection, let us introduce a minimum set of key notions of axiomatic thermodynamics without proof, which are needed for the aim of this paper.

%%%%%%%%%%
\subsubsection{Adiabatic Process and Composition}

\begin{figure}[h]
 \begin{center}
 \includegraphics[height=30mm]{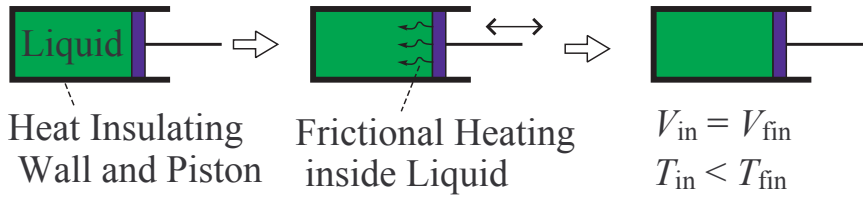}
 \end{center}
\caption{An example of adiabatic process of laboratory system.}
\label{fig:1}
\end{figure}

In an axiomatic formulation of ordinary thermodynamics, for example by Lieb and Ingvason~\cite{ref:lieb+1.1999} or by Tasaki~\cite{ref:tasaki.2000}, the adiabatic process plays the essential role:
\begin{define}[Adiabatic process]
\label{def:adiabatic}
Adiabatic process is the process during which the energy transfer between the system and its environment is given by only mechanical work.
The initial and final states of the system are thermal equilibrium states, but the states during adiabatic process are not necessarily thermal equilibrium states.
\end{define}
An example of adiabatic process is shown in Fig.\ref{fig:1}.
In this example, the system under consideration is a liquid enclosed in heat insulating cylinder and piston.
An adiabatic process is realized by a fast oscillation of piston.
Even when the volume of liquid does not change at initial and final thermal equilibrium states, the frictional heating inside the liquid increases the temperature~\cite{note:principle.1}.
The energy which causes the frictional heating is the mechanical work operated by piston, and hence this process satisfies the definition of adiabatic process.
This adiabatic process is irreversible due to the frictional heating.

It should be emphasized that, in this paper (and in the axiomatic thermodynamics~\cite{ref:lieb+1.1999,ref:tasaki.2000}), the notion of adiabaticity does not mean ``slow''.
The notion of slowness is clearly separated from the notion of adiabaticity, and defined as a {\em quasi-static process} during which not only the initial and final states but also the intermediate states are thermal equilibrium states.
Any quasi-static process (e.g. quasi-static adiabatic process, quasi-static isothermal process, and so on) is reversible.
Note that, in the above example shown in Fig.\ref{fig:1}, the intermediate states of adiabatic process are non-equilibrium states possessing the frictional heating which make the adiabatic process irreversible.

Next, we summarize a useful notion for thermodynamic consideration:
\begin{define}[Composition]
\label{def:composition}
Consider some systems which are individually in thermal equilibrium states, and their thermal equilibrium states are not necessarily the same.
Then, the composition of those systems is simply to regard them as one system.
Each individual system in a composition is called ``subsystem'' of the composite system.
\end{define}
When only one system is under consideration, we may call it {\em the single system} in order to emphasize that we consider only one system and can not consider composition.

Note that, if some subsystems in a composite system interact with each other (e.g. by exchanging heat and/or work), then those subsystems are thermally equilibrium with each other.
However, if a subsystem in a composite system is isolated from the other subsystems, then thermal equilibrium state of the isolated subsystem can be different from equilibrium states of the other subsystems.

It should also be noted that the composition is different from the {\em mixing} in which some systems are mixed into one system (e.g. by removing the wall between two systems).
The mixing is not necessarily needed for understanding the uniqueness of entropy in this paper.

%%%%%%%%%%
\subsubsection{Basic Properties of State Variables}

All state variables in ordinary thermodynamics are distinguished into two categories, {\em extensive variables} and {\em intensive variables}, which are defined by the scaling behavior as follows:
Let $\alpha\,(> 0)$ be a scaling rate of state variables which measure the ``size'' of system, $N \to \alpha\,N$ and $V \to \alpha\,V$, where $N$ is the mol number (number of particles) and $V$ is the volume of the system.
The extensive variable, $X$ (e.g. internal energy and entropy), has the same scaling behavior with the system size, $X \to \alpha\,X$.
The intensive variable, $Y$ (e.g. pressure, temperature and chemical potential), is invariant under the scaling of system size, $Y \to Y$.

Any extensive variable in ordinary thermodynamics is constructed so as to be {\em additive}.
The additivity is expressed as follows:
Consider a composition of $N$ subsystems, and let $X_i$ $(i = 1,2,\cdots,N)$ be an extensive variable (e.g. entropy) of $i$-th subsystem.
Then, the total extensive variable, $X_{\rm com}$, of the composite system is given by
\eqb
\label{eq:td.additivity}
 X_{\rm com} = \sum_{i=1}^{N}X_i \,.
\eqe
This is the additivity of extensive variables.

The above properties of state variables are required as basic principles in axiomatic thermodynamics~\cite{ref:lieb+1.1999,ref:tasaki.2000}.
Then, let us note an important property of state variables derived from the basic principles.
It is the {\em convexity} of various state variables.
Mathematically, a function $f(y)$ is {\em convex}, if $f(\tilde{y}) \le \lambda\,f(y_1) + (1-\lambda)\,f(y_2)$, where $\tilde{y} = \lambda\,y_1 + (1-\lambda)\,y_2$ and $0 < \lambda < 1$.
And, $f(y)$ is {\em concave}, if $-f(y)$ is convex.
(When $f(y)$ is second differentiable, $f(y)$ is convex if ${\rm d}^2\!f(y)/{\rm d}y^2 \ge 0$.)
In ordinary thermodynamics, the convexity of state variables is related with the stability of thermal equilibrium state.
For example, the free energy $F(T,V,N)$ is concave about variables $(T,V,N)$, where $T$ is temperature, $V$ is volume, and $N$ is number of constituent particles.
This yields, for example, the positive heat capacity, $C \defeq - T\,\partial^2\!F(T,V)/\partial T^2 > 0$.
Therefore, the concavity of $F$ implies thermodynamic stability of the system under consideration.

%%%%%%%%%%
\subsubsection{Entropy in Ordinary Thermodynamics}

Finally in this subsection, we review the basic properties of entropy in ordinary thermodynamics.
Entropy is extensive, and therefore it is also additive.

Let us regard the entropy, $S(U,V,N)$, as a function of internal energy $U$, volume $V$ and mol (or particle) number $N$.
Then, it is proven from basic principles of ordinary thermodynamics that $S$ is concave about variables $(U,V,N)$.
Here, for later use, define the entropy density $\sigma(\varepsilon,\rho)$ as
\eqb
\label{eq:td.entropy.density}
 \sigma(\varepsilon,\rho) \defeq \dfrac{S(U,V,N)}{V}
 = S(U/V,1,N/V) = S(\varepsilon,1,\rho) \,,
\eqe
where $\varepsilon \defeq U/V$ is the energy density, and $\rho \defeq N/V$ is the mol density (or number density of constituent particles), and the extensivity of $S$ is used at first equality.
By the definition of concavity,
\eqb
\label{eq:td.entropy.concave}
 \sigma(\tilde{\varepsilon},\tilde{\rho}) \ge
 \lambda\,\sigma(\varepsilon_a,\rho_a) + (1-\lambda)\,\sigma(\varepsilon_b,\rho_b) \,,
\eqe
where $\tilde{\varepsilon} = \lambda\,\varepsilon_a + (1-\lambda)\,\varepsilon_b$~, $\tilde{\rho} = \lambda\,\rho_a + (1-\lambda)\,\rho_b$ and $0<\lambda<1$.
This means $\partial^2\sigma/\partial x^2 \le 0$ ($x = \varepsilon, \rho$), if $\sigma$ is second differentiable.
It is also proven from basic principles of ordinary thermodynamics that $S(U,V,N)$ is monotone increasing about $U$; $\sigma(\varepsilon_a,\rho) < \sigma(\varepsilon_b,\rho)$, where $\varepsilon_a < \varepsilon_b$~.

The notable property of entropy is the theorem called {\em entropy principle} in axiomatic thermodynamics~\cite{ref:lieb+1.1999,ref:tasaki.2000}, which is proven from the basic principles of ordinary thermodynamics~\cite{note:entropy.principle}.
Since a complete review of axiomatic thermodynamics is not the aim of this paper, we show the theorem as one fact of ordinary thermodynamics.
But before showing it, let us introduce a notation of adiabatic process:
Consider two thermal equilibrium states of a single system, and let $(X^{(a)},Y^{(a)})$ be extensive and intensive state variables of one of the two states, and $(X^{(b)},Y^{(b)})$ be those of the other state.
Then, we express the adiabatic process, in which the initial and final states are respectively $(X^{(a)},Y^{(a)})$ and $(X^{(b)},Y^{(b)})$, as~\cite{note:principle.2}
\eqb
\label{eq:td.ad}
 \ad{(X^{(a)},Y^{(a)})}{(X^{(b)},Y^{(b)})} \,,
\eqe
and if this adiabatic process is reversible,
\eqb
\label{eq:td.revad}
 \revad{(X^{(a)},Y^{(a)})}{(X^{(b)},Y^{(b)})} \,.
\eqe
Furthermore, if the system is a composition of $N$ subsystems and $(X_i,Y_i)$ is state variables of $i$-th subsystem ($i=1,\cdots,N$), then we express the adiabatic process~\eref{eq:td.ad} of this composite system as
\eqb
\label{eq:fact1.ad}
 \ad{\{\,(X_1^{(a)},Y_1^{(a)}),\cdots,(X_N^{(a)},Y_N^{(a)}) \,\}}
    {\{\,(X_1^{(b)},Y_1^{(b)}),\cdots,(X_N^{(b)},Y_N^{(b)}) \,\}} \,,
\eqe
and also ${\mathcal Ad_{rev}}$ similarly.
Using these notations, the fact of ordinary thermodynamics is:
\begin{fact}[Entropy principle]
\label{fact:entropy}
Consider a composition of $N$ subsystems.
Let $(X_i,Y_i)$ be the extensive and intensive variables of each subsystem, and $S_{\rm com} = \sum_{i=1}^{N}S_i$ be the total entropy of the composite system.
Then, the necessary and sufficient condition for the existence of an adiabatic process~\eref{eq:fact1.ad} is that the following inequality of total entropy holds,
\eqb
\label{eq:fact1.entropy}
 \sum_{i=1}^{N} S_i^{(a)} \,\le\, \sum_{i=1}^{N} S_i^{(b)} \,.
\eqe
The equality, $S_{\rm com}^{(a)} = S_{\rm com}^{(b)}$, holds if and only if the adiabatic process in Eq.\eref{eq:fact1.ad} is reversible, ${\mathcal Ad_{rev}}$.
\end{fact}
Note that this fact is sometimes regarded as the statement of the second law of thermodynamics.
The identification of entropy principle with the second low of thermodynamics (e.g. the Kelvin's statement of it) is good in rough sense.
However, rigorously speaking, the entropy principle is not equivalent to, for example, the Kelvin's statement of second low, because some basic principles other than the Kelvin's second low are necessary to derive the entropy principle~\cite{ref:lieb+1.1999,ref:tasaki.2000,note:entropy.principle}.

The entropy principle clarifies the thermodynamic meaning of entropy, how the ``direction'' of adiabatic process is determined.
A significant example is as follows:
Let, as an example, $(V_1,T_1)$ be the volume and temperature of a single system named ``1'', and $S_1$ be the entropy of this system.
By the entropy principle, the adiabatic process, $\ad{(V_1^{(a)},T_1^{(a)})}{(V_1^{(b)},T_1^{(b)})}$, is impossible if $S_1^{(a)} > S_1^{(b)}$.
However, construct a composite system with another system ``2'' of state variables $(V_2,T_2)$, and let the subsystems ``1'' and ``2'' interact thermodynamically with each other.
Then, an adiabatic process, $\ad{\{(V_1^{(a)},T_1^{(a)}),(V_2^{(a)},T_2^{(a)})\}}{\{(V_1^{(b)},T_1^{(b)}),(V_2^{(b)},T_2^{(b)})\}}$, becomes possible, if $S_1^{(a)} + S_2^{(a)} < S_1^{(b)} + S_2^{(b)}$ even when $S_1^{(a)} > S_1^{(b)}$.
This denotes that the impossible adiabatic change of state variables of a single system, $(V_1^{(a)},T_1^{a}) \rightsquigarrow (V_1^{(b)},T_1^{(b)})$, can be realized as a part of adiabatic process of an appropriate composite system, if the composition is possible so that $S_{\rm com}^{(a)} < S_{\rm com}^{(b)}$.

By the entropy principle together with the extensivity and additivity of entropy, the uniqueness of entropy is proven in axiomatic thermodynamic~\cite{ref:lieb+1.1999,ref:tasaki.2000}.
However in black hole thermodynamics, as explained below, the extensivity/intensivity classification of state variables (i.e. the scaling behavior of state variables) is modified to some other classification, and the additivity should be re-considered.
Hence, the uniqueness of Bekenstein-Hawking entropy is not manifest in black hole thermodynamics.

%%%%%%%%%%%%%%%%%%%%%%%%%%%%%%%%%%%%%%%%%%%%%%%%%%
\subsection{Black Hole Thermodynamics}
\label{sec:td.bh}

This subsection formulates black hole thermodynamics without using any existing model of quantum gravity.
Planck units are used throughout in this subsection, $c = 1$ , $G = 1$ , $\hbar = 1$ , $k_B = 1$.

%%%%%%%%%%
\subsubsection{Thermal Equilibrium of Schwarzschild Black Hole}

The theoretical basis for regarding a black hole as a thermal equilibrium state of gravitational field is given by the quantum field theory on black hole spacetime, which concludes that any matter field is radiated from the black hole horizon with thermal spectrum ({\em Hawking radiation})~\cite{ref:birrell+1.1982,ref:hawking.1975}.
The suitable situation for considering the black hole thermodynamic is shown in Fig.2:
Enclose a single black hole in a concentric spherical cavity.
Adjust the temperature of heat bath to that determined by the thermal spectrum of Hawking radiation.
Then, the energy coming from the black hole to surface of heat bath due to the Hawking radiation, which is absorbed by the heat bath, balances completely with the energy coming from the heat bath to black hole due to the thermal radiation emitted by heat bath.
Thus, the two-component system, which consists of the black hole and radiation in cavity, is in a thermal equilibrium state of temperature of Hawking radiation.

\begin{figure}[h]
 \begin{center}
 \includegraphics[height=30mm]{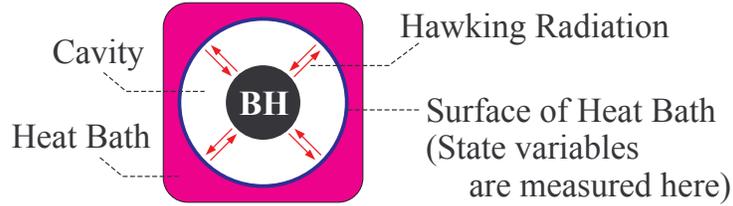}
 \end{center}
\caption{A schematic image of thermal equilibrium state of single black hole (BH).}
\label{fig:2}
\end{figure}

It should be noted that the temperature of Hawking radiation is extremely lower than the mass energy of black hole when the mass is greater than Planck mass~\cite{note:temp/mass}.
Therefore, in calculating state variables of thermal system shown in Fig.\ref{fig:2}, it is physically reasonable to ignore the thermal radiation in cavity.
Such a calculation of state variables in black hole thermodynamics is carried out, for the first, by York in the framework of Euclidean quantum gravity~\cite{ref:braden+3.1990,ref:brown+2.1991,ref:york.1986}.
However, we modify the York's discussion so as to construct the black hole thermodynamics without using any existing model of quantum gravity:

Let the observer be at the surface of heat bath, and the areal radius of the surface be $r_w$.
This means that state variables of black hole are measured at $r = r_w$.
Consider, for simplicity, a Schwarzschild black hole of mass $M$ (horizon radius $2 M$), whose metric in Schwarzschild coordinates $(t,r,\theta,\phi)$ is given by the line element of spacetime,
\eqb
\label{eq:td.metric}
 ds^2 = - \left(1-\dfrac{2M}{r}\right)\,dt^2 + \dfrac{1}{1-2M/r}\,dr^2
 + r^2\,\left(\, d\theta^2 + \sin^2\theta\,d\phi^2 \,\right) \,,
\eqe
where $2M < r_w$ should hold in order to let this black hole be in the cavity.
Note two points:
First point is that, because the region connected causally to the observe is the region between the event horizon and surface of heat bath, $2M < r < r_w$, the black hole thermodynamics should be described in that region.
Therefore, the Schwarzschild coordinate is suitable for black hole thermodynamics, because the line element~\eref{eq:td.metric} is expressed in the static form in that region~\cite{note:equilibrium}.
Second point is that the number of independent state variables of Schwarzschild black hole is two, due to the two parameters, $M$ and $r_w$.

Then, as the state variable of size of the system in Fig.\ref{fig:2}, one may consider a proper three volume of cavity, $\int_{2M}^{r_w} dr\,4\pi\,r^2/\sqrt{1-2M/r}$ (this integral converges).
However, as explained later in this subsection, any three volume can not produce a consistent scaling behavior in black hole thermodynamics.
Hence, we can not adopt three volume as state variable of system size, but it is known that the consistent state variable of system size is the area at surface of heat bath~\cite{ref:york.1986},
\eqb
\label{eq:td.Aw}
 A_w = 4\pi r_w^2 \,.
\eqe
This $A_w$ is measurable at $r_w$ and has the same scaling behavior with the other extensive variables (e.g. entropy) as explained later in this subsection.

Next, the temperature of black hole can be read from thermal spectrum of Hawking radiation~\cite{ref:hawking.1975,ref:york.1986},
\eqb
\label{eq:td.Tbh}
 T_{\rm BH} = \gamma_{\rm tol}\,\dfrac{\kappa}{2\pi} \,,
\eqe
where $\kappa = 1/4M$ is the surface gravity of black hole, and $\gamma_{\rm tol} = 1/\sqrt{1-2M/r_w}$ is called the Tolman factor which expresses the gravitational redshift received by the Hawking radiation during propagating from black hole horizon to surface of heat bath~\cite{ref:tolman.1987}.

The free energy of black hole may be usually calculated in the framework of Euclidean quantum gravity~\cite{ref:gibbons+1.1977,ref:hawking.1979,ref:york.1986}.
However, we have to emphasize that the Euclidean quantum gravity is not the only method for obtaining the free energy of black hole. 
As shown in App.\ref{app:free} in detail, the free energy of Schwarzschild black hole can be constructed (without the Euclidean quantum gravity) by accepting two requirements that the entropy is given by Eq.\eref{eq:td.Sbh} and that the asymptotic value of internal energy as $r_w\to\infty$ is $M$. 
The physically natural reason for adopting these two requirements is explained in App.\ref{app:free} in detail, which is based on general relativity, quantum field theory and ordinary thermodynamics. 
(That physical reasoning forms the basis of the notion of black hole thermodynamics.)
The resultant form of free energy is
\eqb
\label{eq:td.Fbh}
 F_{\rm BH} =
  r_w\,\left(\,1-\sqrt{1-\dfrac{2M}{r_w}}\,\right)
 - \dfrac{M}{2\sqrt{1-2M/r_w}} \,.
\eqe
(York has obtained the same free energy with Eq.\eref{eq:td.Fbh} in the framework of Euclidean quantum gravity~\cite{ref:york.1986}.)
Adopting the construction of $F_{\rm BH}$ as given in App.\ref{app:free}, all calculations in this paper become independent of details of existing quantum gravity models.

Given the above three state variables, the other state variables are obtained by following the procedure of ordinary thermodynamics.
The Bekenstein-Hawking entropy is given by
\eqb
\label{eq:td.Sbh}
 S_{\rm BH} =
 -\pd{F_{\rm BH}(T_{\rm BH},A_w)}{T_{\rm BH}} = \pi\,(2M)^2 \,,
\eqe
where, following ordinary thermodynamics, the free energy is regarded as a function of temperature and system size. 
(See App.\ref{app:free}, in which the role of Eq.\eref{eq:td.Sbh} in constructing $F_{\rm BH}$ is explained.) 
Given the entropy, the heat capacity of black hole is calculated,
\eqb
\label{eq:td.Cbh}
 C_{\rm BH} \defeq T_{\rm BH}\,\pd{S_{\rm BH}(T_{\rm BH},A_w)}{T_{\rm BH}}
 = -8\pi M^2\,\dfrac{1-2M/r_w}{1-3M/r_w} \,.
\eqe
Next, the internal energy is given by the Legendre transformation,
\eqb
\label{eq:td.Ubh}
 U_{\rm BH} \defeq F_{\rm BH} + S_{\rm BH}\,T_{\rm BH}
 = r_w\,\left(\,1-\sqrt{1-\dfrac{2 M}{r_w}} \,\right) \,.
\eqe
(See App.\ref{app:free}, in which the role of the limit, $\lim_{r_w\to\infty}=M$, in constructing $F_{\rm BH}$ is explained.) 
The state variable which is thermodynamically conjugate to $A_w$ is given by
\eqb
\label{eq:td.Pw}
 P_w \defeq -\pd{U_{\rm BH}(S_{\rm BH},A_w)}{A_w}
 = \dfrac{1}{8 \pi r_w}\,\left(\,\dfrac{1-M/r_w}{\sqrt{1-2M/r_w}} - 1\,\right) \,.
\eqe
This $P_w$ corresponds to the pressure in the ordinary gas in laboratory, but the dimension of $P_w$ is not of the pressure.
The detail of thermodynamic meaning of $P_w$ is explained in appendix~B of~\cite{ref:saida.2009a}, but the detail is not necessarily needed for the aim of this paper.
The point is that state variables of the system shown in Fig.\ref{fig:2} can be defined independently of details of existing quantum gravity models.

%%%%%%%%%%
\subsubsection{Thermal Stability of Schwarzschild Black Hole}

\begin{figure}[h]
 \begin{center}
 \includegraphics[height=45mm]{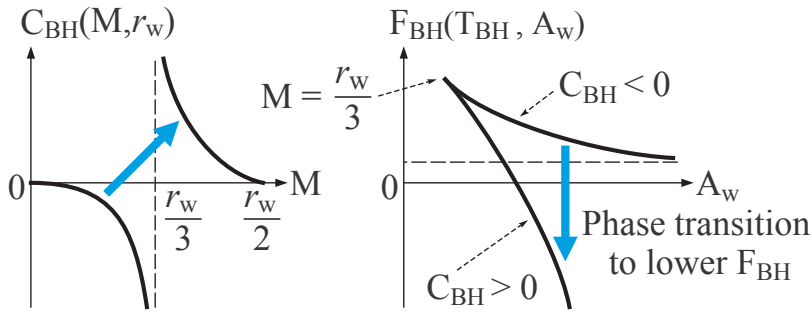}
 \end{center}
\caption{Schematic graphs of $C_{\rm BH}(M,r_w)$ and $F_{\rm BH}(T_{\rm BH},A_w)$.}
\label{fig:3}
\end{figure}

We should specify the range of parameters, $M$ and $r_w$, so as to let the thermal equilibrium state of black hole be thermodynamically stable.
Fig.\ref{fig:3} shows schematic graphs of heat capacity $C_{\rm BH}$ as function of $M$ and $r_w$, and free energy $F_{\rm BH}$ as function of $T_{\rm BH}$ and $A_w$.
We find that the thermal equilibrium state of black hole is thermally unstable for $3M < r_w$ due to negative heat capacity, while it is thermally stable for $2M < r_w < 3M$ due to positive heat capacity.
From solely the behavior of heat capacity, one may think that black hole thermodynamics is ill-defined, since thermal stability is not necessarily ensured.
However, if we assume that the criterion of phase transition in ordinary thermodynamics is applicable to black hole, it is concluded from the behavior of $F_{\rm BH}$ shown in Fig.\ref{fig:3} that an unstable equilibrium state is transformed to a stable one under the environment of constant temperature, because $F_{\rm BH}(2M<r_w<3M) < F_{\rm BH}(3M,r_w)$.
This denotes that a consistent thermodynamic formulation of thermal system shown in Fig.\ref{fig:2} is expected for the range of parameters,
\eqb
\label{eq:td.range}
 2M < r_w < 3M \,.
\eqe
The same evidence, that this parameter region is suitable for black hole thermodynamics, is also obtained by considering a mechanical stability based on the positivity of isentropic compressibility defined by $A_w^{-1}(\partial A_w(S_{\rm BH},P_w)/\partial P_w)$, which is considered for the first by York~\cite{ref:york.1986} and rigorously defined in appendix~B of~\cite{ref:saida.2009a}.
In the following part of this subsection, we show the evidence that a consistent Schwarzschild black hole thermodynamics can be constructed for this parameter range.

%%%%%%%%%%
\subsubsection{Scaling Behavior of State Variables}

In order to discuss the classification of state variables, recall that, in ordinary thermodynamics for laboratory systems, the extensive and intensive variables are defined via the scaling behavior of the size of system; volume and mol number.
However, in Schwarzschild black hole thermodynamics, the fundamental parameters, $M$ and $r_w$, have the dimension of length, not of mol number.
Thus, the fundamental scaling should be of the length scaling as shown below in Eq.\eref{eq:td.scaling}.
Then, as implied by Eqs.\eref{eq:td.Aw}--\eref{eq:td.Pw}, we define the classification of state variables in black hole thermodynamics as follows:
\begin{define}[Classification of state variables of black hole]
\label{def:classification}
Let the fundamental scaling be
\eqb
\label{eq:td.scaling}
 M \to \lambda\,M \quad,\quad r_w \to \lambda\,r_w \,,
\eqe
where $\lambda\,\,(>0)$ is the rate of scaling of ``length size''.
Under this fundamental scaling, all state variables in black hole thermodynamics are classified into three categories:
\begin{description}
\item[\,\,Extensive variable:]
These variables, $X$ (e.g. $A_w$, $S_{\rm BH}$ and $C_{\rm BH}$), are scaled as, $X \to \lambda^2\,X$.
\item[\,\,Intensive variable:]
These variables, $Y$ (e.g. $T_{\rm BH}$ and $P_w$), are scaled as, $Y \to \dfrac{Y}{\lambda}$.
\item[\,\,Thermodynamic energy:]
These energies, $Z$ (e.g. $F_{\rm BH}$ and $U_{\rm BH}$), are scaled as, $Z \to \lambda\,Z$.
\end{description}
Here, thermodynamic energy is the state variable possessing the dimension of energy and related with free energy by the Legendre transformation.
\end{define}
The same classification of state variables is also found in the other black hole thermodynamics; Reissner-Nortstr\"{o}m black hole~\cite{ref:braden+3.1990}, Kerr black hole~\cite{ref:brown+2.1991} and so on~\cite{ref:saida.2009a,ref:saida.2009b}.

This classification is one of different points of black hole thermodynamics from ordinary thermodynamics.
The proof of uniqueness of Bekenstein-Hawking entropy should be constructed with the above scaling behavior of black hole thermodynamics.

%%%%%%%%%%
\subsubsection{Adiabatic Process and Composition}

In the proof of uniqueness of Bekenstein-Hawking entropy, the adiabatic process and composition of some systems are used.
The definition of them are the same as given in Subsec.\ref{sec:td.axiom}.
In order to understand those notions in the context of black hole thermodynamics, let us show an example of adiabatic process, and make comments on the composition of some systems.

\begin{figure}[h]
 \begin{center}
 \includegraphics[height=45mm]{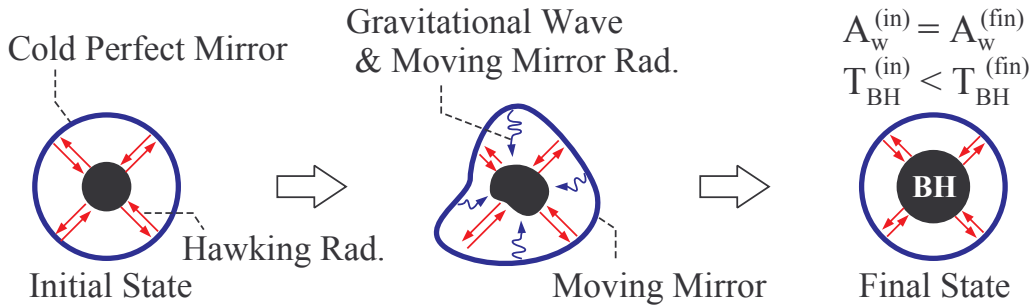}
 \end{center}
\caption{An example of adiabatic process in black hole thermodynamics.}
\label{fig:4}
\end{figure}

Fig.\ref{fig:4} shows an example of adiabatic process of black hole which corresponds to the adiabatic process of laboratory system shown in Fig.\ref{fig:1}:
The ``heat insulating'' environment in black hole thermodynamics, which plays a role of heat insulating wall in laboratory system, is a perfectly reflecting mirror of zero temperature ({\em the cold perfect mirror}).
Even when the (surface of) heat bath of thermal equilibrium system shown in Fig.\ref{fig:2} is replaced with the cold perfect mirror, the black hole is still in a thermal equilibrium state, because the Hawking radiation is perfectly reflected at the cold perfect mirror and the energy balance is realized between the Hawking radiation and reflected radiation.
Here note that, if the mirror has some finite temperature, the thermal radiation due to the temperature violates the energy balance, and the state of system becomes non-equilibrium.
Therefore, the zero temperature of mirror is necessary.

For the process shown in Fig.\ref{fig:4}, let the initial and final equilibrium states have the same system size, $A_w^{\rm (in)} = A_w^{\rm (fin)}$, and the intermediate states be non-equilibrium states as follows.
Suppose that the shape of cold perfect mirror is deformed dynamically by some mechanical work, and the mirror keep moving during the process.
By this moving mirror, there arise classical and quantum effects.
The classical effect is the radiation of gravitational wave due to asymmetric motion of the mirror.
By the argument of general relativity, the energy of black hole does not decrease by this classical effect.
The quantum effect is the moving mirror radiation created due to the motion of mirror, at which a boundary condition is imposed on quantum fields~\cite{ref:birrell+1.1982,ref:fulling+1.1976}.
The quantum radiation by moving mirror is analogous to Hawking radiation in the sense that the time evolution of boundary condition of quantum fields changes the quantum vacuum state and creates quantum particles which constitute the radiation~\cite{ref:birrell+1.1982}.
Because the moving mirror radiation injects an energy from mirror to black hole, the energy of black hole increases also by this effect.
Therefore, during the process shown in Fig.\ref{fig:4}, the black hole is non-stationary and increases its energy due to the classical and quantum radiation by moving mirror.
Such a radiation by moving mirror corresponds to the frictional heating in the adiabatic process of laboratory system shown in Fig.\ref{fig:1}.
Furthermore, by Eqs.\eref{eq:td.Tbh} and \eref{eq:td.Ubh}, it is found that, within the parameter range~\eref{eq:td.range}, the increase of thermodynamic energy $U_{\rm BH}$ due to the effects of moving mirror results in the increase of mass $M$, and the increase of $M$ causes the increase of $T_{\rm BH}$.
Hence, the black hole temperature increases in the adiabatic process shown in Fig.\ref{fig:4} as that of laboratory system shown in Fig.\ref{fig:1}.

\begin{figure}[h]
 \begin{center}
 \includegraphics[height=60mm]{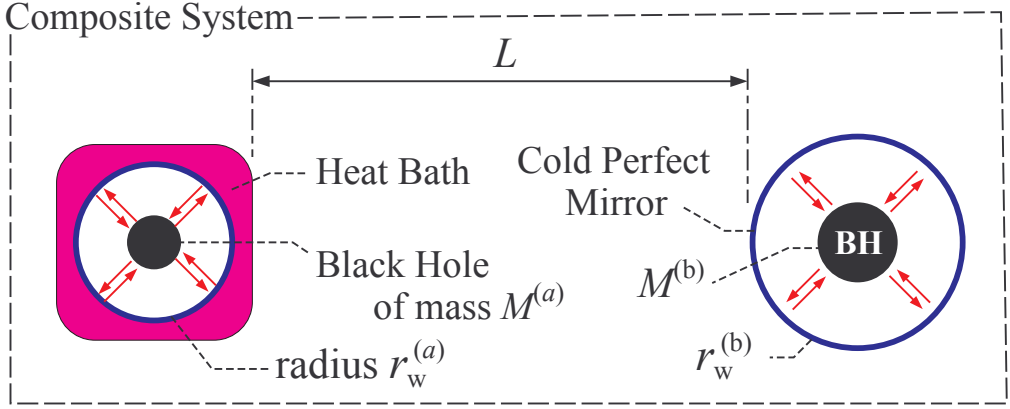}
 \end{center}
\caption{An example of composition of two equilibrium systems of black holes.}
\label{fig:5}
\end{figure}

Next, we make comments on the composition of some systems in black hole thermodynamics.
Fig.\ref{fig:5} shows an example of composition, in which subsystems are two thermal systems of black holes.
(The ``wall'' of each subsystem, which encloses a black hole, can be either heat bath or cold perfect mirror.)
These two subsystems are individually in thermal equilibrium states by equilibrating each black hole with the heat bath or cold perfect mirror.

We have three comments related with the composite system.
First comment is on the additivity of Bekenstein-Hawking entropy, which is used in the proof of its uniqueness.
Note that it is already revealed in~\cite{ref:martinez+1.1989} that the total entropy of a two-component system of a black hole and a matter field, such as the system shown in Fig.\ref{fig:2}, satisfies the additivity, $S_{\rm tot} = S_{\rm BH} + S_{\rm matter}$, where $S_{\rm matter}$ is matter entropy.
Thus, it may be reasonable to require also the additivity of entropy in the multi-black hole composite system as shown in Fig.\ref{fig:5},
\eqb
\label{eq:td.additive}
 S_{\rm com} = S_{\rm BH}^{(a)} + S_{\rm BH}^{(b)} \,.
\eqe
Rigorously speaking, this additivity is simply an assumption for the case of short separation length, $L$, between the subsystems.
However, if the separation length is so large, $L \gg M^{(a)},\,M^{(b)}$, that the gravitational potential between these subsystems is much less than the mass of black holes, $M^{(a)}$ and $M^{(b)}$, then the validity of additivity~\eref{eq:td.additive} is obvious.

The second comment is on the concrete form of Bekenstein-Hawking entropy in the composite system.
Note that, it is already revealed in~\cite{ref:saida.2009b} that, even when the subsystems in Fig.\ref{fig:5} are individually in thermal equilibrium states, the Bekenstein-Hawking entropy is not necessarily expressed by the entropy-area law~\eref{eq:td.Sbh} unless the gravitational interaction between these subsystems are ignored.
(The entropy-area law is applicable only to single black hole which is not affected by the gravity of other horizons.)
Thus, when we need the entropy-area law~\eref{eq:td.Sbh} as the Bekenstein-Hawking entropy even for subsystems in a composite system, we should make the separation length between subsystems, $L$, be so long that the gravitational potential between these subsystems is much less than the mass of black holes.
With such a long $L$, the entropy-area law~\eref{eq:td.Sbh} becomes applicable to the entropy of each black hole.
However, the entropy-area law is not needed in the proof of uniqueness of Bekenstein-Hawking entropy.
Thus, we can let $L$ be arbitrary if the additivity~\eref{eq:td.additive} holds.

The third comment is on the additivity of state variables other than entropy.
There is no reason to deny the additivity of system size for the composite system, $A_{\rm com} = A_w^{(a)} + A_w^{(b)}$.
On the other hand, it is already revealed in~\cite{ref:martinez+1.1989} that thermodynamic energy is not additive for the two-component system of a black hole and a matter field.
{\em Hence, under the classification of state variables in definition~\ref{def:classification}, it is physically reasonable to require that the extensive variable is additive also in black hole thermodynamics, however the thermodynamic energy, which is an extensive variable in ordinary thermodynamics but not in black hole thermodynamics, becomes non-additive in black hole thermodynamics.}

%%%%%%%%%%
\subsubsection{Basic Properties of Bekenstein-Hawking Entropy}

There are six basic properties of Bekenstein-Hawking entropy, necessary for the aim of this paper:
\begin{itemize}
\item
Extensivity of $S_{\rm BH}$ formulated in definition~\ref{def:classification}.
\item
Additivity of $S_{\rm BH}$ in Eq.\eref{eq:td.additive}.
\item
Entropy principle shown in fact~\ref{fact:entropy}. (See comments given below.)
\item
Uniqueness of $S_{\rm BH}$. (Proof is given in the next subsection.)
\item
Concavity of $S_{\rm BH}$ about internal energy $U_{\rm BH}$ and system size $A_w$. (Detail is given below.)
\item
Monotone increasing nature of $S_{\rm BH}$ about $U_{\rm BH}$. (Detail is given below.)
\end{itemize}
First three of these properties (extensivity, additivity and entropy principle) are used in the proof of uniqueness of Bekenstein-Hawking entropy.
The rest of these properties (uniqueness, concavity and monotone increasing nature) are referred to in obtaining the conclusion of this paper.

Extensivity and additivity are already explained.
Here, let us show the concavity, monotone increasing nature and entropy principle:
The concavity and monotone increasing nature, for Schwarzschild black hole, can be explicitely obtained from Eqs.\eref{eq:td.Aw}, \eref{eq:td.Sbh} and~\eref{eq:td.Ubh} as follows.
Bekenstein-Hawking entropy is rearranged to, $S_{\rm BH}(U_{\rm BH},A_w) = 4\pi\,[\, U_{\rm BH} - U_{\rm BH}^2\sqrt{\pi/A_w}\,]^2$.
Then, it is straightforward to show inequalities,
\eqb
 \pd{S_{\rm BH}(U_{\rm BH},A_w)}{U_{\rm BH}} > 0 \quad,\quad
 \pd{^2 S_{\rm BH}(U_{\rm BH},A_w)}{U_{\rm BH}^2} < 0 \quad,\quad
 \pd{^2 S_{\rm BH}(U_{\rm BH},A_w)}{A_w^2} <~0 \,,
\eqe
for the parameter range~\eref{eq:td.range}.
The first inequality denotes that $S_{\rm BH}(U_{\rm BH},A_w)$ is monotone increasing about $U_{\rm BH}$, and the last two inequalities denote that $S_{\rm BH}(U_{\rm BH},A_w)$ is concave about $(U_{\rm BH},A_w)$

Next, we comment on the entropy principle in black hole thermodynamics.
Following the axiomatic formulation of ordinary thermodynamics~\cite{ref:lieb+1.1999,ref:tasaki.2000} together with appropriate modifications due to the peculiar classification of state variables shown in definition~\ref{def:classification}, it is possible to prove that the entropy principle, given in fact~\ref{fact:entropy}, holds also in black hole thermodynamics.
However, the proof of entropy principle is too lengthy so that it would make the paper twice (or more) as length as present paper. 
{\em Thus, in this paper, please let us accept the entropy principle as a fact also in black hole thermodynamics with a physical reasoning as follows}:

For a physical reasoning of accepting the entropy principle in black hole thermodynamics, recall the comment given just below the fact~\ref{fact:entropy} that the identification of entropy principle with the second law of thermodynamics is good in rough sense.
On the other hand, in black hole thermodynamics, the validity of so-called {\em generalized second law} is widely checked with various processes of systems including black hole~\cite{ref:bekenstein.1974,ref:flanagan+2.2000,ref:frolov+1.1993,ref:saida.2006,ref:unruh+1.1982,ref:unruh+1.1983}.
Thus, we can find a physical understanding that the entropy principle in black hole thermodynamics is justified by the validity of generalized second law.
(The proof of entropy principle and axiomatic formulation of black hole thermodynamics may appear in the other paper in future.)

%%%%%%%%%%%%%%%%%%%%%%%%%%%%%%%%%%%%%%%%%%%%%%%%%%
\subsection{Uniqueness Theorem of Bekenstein-Hawking Entropy}

\subsubsection{Statement of Theorem}

\begin{thm}[Uniqueness of Bekenstein-Hawking entropy]
\label{thm:uniqueness}
Consider the thermal equilibrium system of black hole satisfying all the above preparations.
If there exists a state variable, $K_{\rm BH}$, satisfying the extensivity, additivity and entropy principle with replacing $S_{\rm BH}$ with $K_{\rm BH}$ in the statement of fact~\ref{fig:1}, then $K_{\rm BH}$ is equivalent to the Bekenstein-Hawking entropy, $S_{\rm BH}$, in the sense that $K_{\rm BH}$ is an affine transformation of $S_{\rm BH}$,
\eqb
\label{eq:td.uniqueness}
 K_{\rm BH} = \alpha\,S_{\rm BH} + \eta \,,
\eqe
where $\alpha\,(>0)$ is a positive constant and $\eta$ is a constant satisfying the scaling behavior of extensive variable. 
This is the uniqueness of Bekenstein-Hawking entropy.
\end{thm}

As shown below, the concavity and monotone increasing nature of $S_{\rm BH}$, which are shown from Eq.\eref{eq:td.Sbh} in Subsec.\ref{sec:td.bh}, are not used in the proof of this theorem.
Therefore, the entropy-area law~\eref{eq:td.Sbh} is not necessary for this theorem, once the additivity, extensivity and entropy principle are accepted.

\subsubsection{Proof of Theorem~\ref{thm:uniqueness} (preparations)}

Our proof follows that given in Tasaki's book~\cite{ref:tasaki.2000} with modifications due to the scaling behavior in definition~\ref{def:classification}.

For the clarity of statements, express the Bekenstein-Hawking entropy as $S_{\rm BH}(X,Y;Z)$, where
$X$ is extensive variable, $Y$ intensive variable, $Z$ thermodynamic energy.
We choose, for example, the first two arguments $X$ and $Y$ as the independent state variables.
Then, the third argument, $Z(X,Y)$, becomes dependent one.
Also, express $K_{\rm BH}$ as $K_{\rm BH}(X,Y;Z)$.
(Since two state variables are independent in thermal system of single Schwarzschild black hole, it is redundant to show the dependent variable in the argument of $S_{\rm BH}(X,Y;Z)$ and $K_{\rm BH}(X,Y;Z)$.
However, we do so in order to express clearly the scaling behavior of all three categories of state variables given in definition~\ref{def:classification}.)

Introduce two fixed values of extensive variable, $X^{(1)}$ and $X^{(2)}$ (e.g. the system size $A_w^{(i)}$, $i = 1,\,2$), and one fixed value of intensive variable, $Y^{(0)}$ (e.g. the temperature $T_{\rm BH}^{(0)}$), so that the inequality holds,
\eqb
\label{eq:td.proof.pre.1}
 S_{\rm BH}(X^{(1)},Y^{(0)};Z^{(1,0)}) < S_{\rm BH}(X^{(2)},Y^{(0)};Z^{(2,0)}) \,,
\eqe
where $Z^{(i,0)} \defeq Z(X^{(i)},Y^{(0)})$, $i = 1,\,2$.
Next, determine two constants, $\alpha$ and $\eta$, by two algebraic equations,
\eqb
\label{eq:td.proof.pre.2}
 K_{\rm BH}(X^{(i)},Y^{(0)};Z^{(i,0)})
  = \alpha\,S_{\rm BH}(X^{(i)},Y^{(0)};Z^{(i,0)}) + \eta \quad,\quad
 i = 1,\,2 \,.
\eqe
Then, the proof consists of following four steps:
\begin{description}
\item[\,\,\,Step 1:] Show the positivity of $\alpha > 0$.
\item[\,\,\,Step 2:] Show that the relation, $K_{\rm BH}(X,Y^{(0)};Z^{(0)}) = \alpha\,S_{\rm BH}(X,Y^{(0)};Z^{(0)}) + \eta$, holds with arbitrary $X$, where $Z^{(0)} \defeq Z(X,Y^{(0)})$.
\item[\,\,\,Step 3:] Show that the relation, $K_{\rm BH}(X,Y;Z) = \alpha\,S_{\rm BH}(X,Y;Z) + \eta$, holds with arbitrary $Y$.
\item[\,\,\,Step 4:] Show the extensivity of $\eta$.
\end{description}

%%%%%%%%%%
\subsubsection{Step 1 of the Proof}

By the entropy principle of $S_{\rm BH}$, Eq.\eref{eq:td.proof.pre.1} denotes the existence of an irreversible adiabatic process, $\ad{(X^{(1)},Y^{(0)};Z^{(1,0)})}{(X^{(2)},Y^{(0)};Z^{(2,0)})}$.
Then, by the presupposition that $K_{\rm BH}$ satisfies the entropy principle, we find an inequality, $K_{\rm BH}(X^{(1)},Y^{(0)};Z^{(1,0)}) < K_{\rm BH}(X^{(2)},Y^{(0)};Z^{(2,0)})$.
Therefore, by definition of $\alpha$ given in Eq.\eref{eq:td.proof.pre.2}, the step~1 ends,
\eqb
 \alpha =
 \dfrac{K_{\rm BH}(X^{(2)},Y^{(0)};Z^{(2,0)}) - K_{\rm BH}(X^{(1)},Y^{(0)};Z^{(1,0)})}
       {S_{\rm BH}(X^{(2)},Y^{(0)};Z^{(2,0)}) - S_{\rm BH}(X^{(1)},Y^{(0)};Z^{(1,0)})}
 > 0 \,.
\eqe

%%%%%%%%%%
\subsubsection{Step 2 of the Proof}

For arbitrary extensive variable, $X$, there are two possible cases of entropy;
\eqb
\label{eq:td.proof.2.0}
 \begin{cases}
  \mbox{case (a)} &:\quad S_{\rm BH}(X,Y^{(0)};Z^{(0)}) \le S_{\rm BH}(X^{(1)},Y^{(0)};Z^{(1,0)}) \\
  \mbox{case (b)} &:\quad S_{\rm BH}(X,Y^{(0)};Z^{(0)}) > S_{\rm BH}(X^{(1)},Y^{(0)};Z^{(1,0)})
 \end{cases} \,.
\eqe
Let us prove the relation, $K_{\rm BH}(X,Y^{(0)};Z^{(0)}) = \alpha\,S_{\rm BH}(X,Y^{(0)};Z^{(0)}) + \eta$, for each case.

The case (a).
Determine $\lambda_a$ by the algebraic equation,
\eqb
\label{eq:td.proof.2.1}
 \lambda_a^2\,S_{\rm BH}^{(1,0)} + S_{\rm BH}^{(1,0)}
 = \lambda_a^2\,S_{\rm BH}^{(2,0)} + S_{\rm BH}(X,Y^{(0)};Z^{(0)}) \,,
\eqe
where $S_{\rm BH}^{(i,0)} \defeq S_{\rm BH}(X^{(i)},Y^{(0)};Z^{(i,0)})$, $i = 1,\,2$.
By the additivity of Bekenstein-Hawking entropy~\eref{eq:td.additive}, the left- and right-hand sides of Eq.\eref{eq:td.proof.2.1} are, respectively, understood as the total entropy of a composite system composed of two thermal systems of black holes.
Therefore, by the entropy principle of $S_{\rm BH}$ and scaling behavior of state variables in definition~\ref{def:classification}, Eq.\eref{eq:td.proof.2.1} denotes the existence of a reversible adiabatic process of the composite system,
\begin{multline}
 \revad
 {\Bigl\{\,\Bigl(\,\lambda_a^2\,X^{(1)}\,,\,\dfrac{Y^{(0)}}{\lambda_a}
               \,,\,\lambda_a\,Z^{(1,0)}\,\Bigr)\,,\,
          (X^{(1)}\,,\,Y^{(0)}\,,\,Z^{(1,0)})\,\Bigr\}\\}
 {\Bigl\{\,\Bigl(\lambda_a^2\,X^{(2)}\,,\,\dfrac{Y^{(0)}}{\lambda_a}
               \,,\,\lambda_a\,Z^{(2,0)}\Bigr)\,,\,
          (X\,,\,Y^{(0)}\,,\,Z^{(0)})\,\Bigr\}} \,.
\end{multline}
Then, by the presupposition that $K_{\rm BH}$ is extensive and additive, and satisfies the entropy principle, this reversible adiabatic process denotes that the following relation holds,
\eqb
\label{eq:td.proof.2.2}
 \lambda_a^2\,K_{\rm BH}^{(1,0)} + K_{\rm BH}^{(1,0)}
 = \lambda_a^2\,K_{\rm BH}^{(2,0)} + K_{\rm BH}(X,Y^{(0)};Z^{(0)}) \,,
\eqe
where $K_{\rm BH}^{(i,0)} \defeq K_{\rm BH}(X^{(i)},Y^{(0)};Z^{(i,0)})$, $i = 1,\,2$.
From Eqs.\eref{eq:td.proof.2.1} and~\eref{eq:td.proof.2.2}, we find,
\eqb
 \lambda_a^2 =\,\,
 \dfrac{S_{\rm BH}^{(1,0)} - S_{\rm BH}(X,Y^{(0)};Z^{(0)})}
       {S_{\rm BH}^{(2,0)} - S_{\rm BH}^{(1,0)}}
 =
 \dfrac{K_{\rm BH}^{(1,0)} - K_{\rm BH}(X,Y^{(0)};Z^{(0)})}
       {K_{\rm BH}^{(2,0)} - K_{\rm BH}^{(1,0)}} \,,
\eqe
where $\lambda_a^2 \ge 0$ holds with the condition~\eref{eq:td.proof.2.0} of case (a).
Substituting Eq.\eref{eq:td.proof.pre.2} into the right-hand side of this relation, the case (a) of step~2 ends,
\eqb
\label{eq:td.proof.2.3}
 K_{\rm BH}(X,Y^{(0)};Z^{(0)}) = \alpha\,S_{\rm BH}(X,Y^{(0)};Z^{(0)}) + \eta \,.
\eqe

Next, the case (b).
Determine $\lambda_b$ by the algebraic equation,
\eqb
 S_{\rm BH}^{(1,0)} + \lambda_b^2\,S_{\rm BH}^{(2,0)}
 = \lambda_b^2\,S_{\rm BH}^{(1,0)} + S_{\rm BH}(X,Y^{(0)};Z^{(0)}) \,.
\eqe
Then, following the same discussion with that in the case (a), we obtain Eq.\eref{eq:td.proof.2.3}.
The step~2 ends.

%%%%%%%%%%
\subsubsection{Step 3 of the Proof}

For arbitrary extensive and intensive variables, $(X,Y)$, consider a reversible adiabatic process,
\eqb
 \revad{\bigl(\,\widetilde{X},Y^{(0)},Z(\widetilde{X},Y^{(0)}) \,\bigr)}
       {\bigl(\,X,Y,Z(X,Y)\,\bigr)} \,,
\eqe
where $\widetilde{X}$ is a value of extensive variable determined by $X$, $Y$, and $Y^{(0)}$ so as to realize this reversible adiabatic process.
By the entropy principle of $S_{\rm BH}$ and $K_{\rm BH}$, this reversible adiabatic process denotes that the following relations hold,
\eqab
\label{eq:td.proof.3.1}
 S_{\rm BH}(\widetilde{X},Y^{(0)};Z(\widetilde{X},Y^{(0)})\,)
  &=& S_{\rm BH}(X,Y;Z(X,Y)\,) \\
\label{eq:td.proof.3.2}
 K_{\rm BH}(\widetilde{X},Y^{(0)};Z(\widetilde{X},Y^{(0)})\,)
  &=& K_{\rm BH}(X,Y;Z(X,Y)\,) \,.
\eqae
By the result of step~2 and Eq.\eref{eq:td.proof.3.2}, we find,
\eqb
 \alpha\,S_{\rm BH}(\widetilde{X},Y^{(0)};Z(\widetilde{X},Y^{(0)})\,) + \eta
 = K_{\rm BH}(X,Y;Z(X,Y)\,) \,.
\eqe
Hence, substituting Eq.\eref{eq:td.proof.3.1} into the left-hand side of this relation, the step~3 ends,
\eqb
 K_{\rm BH}(X,Y;Z(X,Y)\,) = \alpha\,S_{\rm BH}(X,Y;Z(X,Y)\,) + \eta \,.
\eqe

%%%%%%%%%%
\subsubsection{Step 4 of the Proof}

By the result of step~3, we have, $\eta = K_{\rm BH}(X,Y;Z(X,Y)\,) - \alpha\,S_{\rm BH}(X,Y;Z(X,Y)\,)$.
This right-hand side is obviously extensive.
Hence, $\eta$ is also extensive.
The uniqueness theorem is proven. 
$\square$

%%%%%%%%%%%%%%%%%%%%%%%%%%%%%%%%%%%%%%%%%%%%%%%%%%%%%%%%%%%%%%%%%%%%%%%%%%%%%%%%%%%%%%%%%%%%%%%%%%%%
\section{Conditions Justifying Boltzmann formula}
\label{sec:bf}

The essential properties of entropy in ordinary thermodynamics are the entropy principle and uniqueness of entropy.
As shown in Sec.\ref{sec:td}, the Bekenstein-Hawking entropy is also equipped with those essential properties of entropy.
Then, it is reasonable to consider that the Bekenstein-Hawking entropy is calculated, in statistical mechanical sense, by applying the Boltzmann formula to a number of states determined by the underlying quantum gravity.
This implies that the underlying quantum gravity and ordinary quantum mechanics share the same properties which justify the Boltzmann formula.

The aim of this section is to show the intrinsic properties of quantum mechanics which justify the Boltzmann formula.
Thus, this section does not refer to the general relativity and black hole thermodynamics.
A reader, who knows a Dobrushin theorem in~\cite{ref:dobrushin.1964} and the chapter~3 of Ruelle's book~\cite{ref:ruelle.1999} which is also found in the appendix~C of Tasaki's book~\cite{ref:tasaki.2008}, can skip this section.
Other reader, who needs only the statements of main theorems without proof, see only Subsec.\ref{sec:bf.thm}.
Relation of the contents of this section with black hole thermodynamics is discussed in the next section.
We use the units, $\hbar = 1$ and $k_B = 1$, in this section.

%%%%%%%%%%%%%%%%%%%%%%%%%%%%%%%%%%%%%%%%%%%%%%%%%%
\subsection{Statements of Theorems and a Corollary without Proof}
\label{sec:bf.thm}

Let us start this subsection with summarizing the basic setting and notations.
Consider a non-relativistic quantum system, and let the system be made of identical particles, for simplicity.
Let $V$ denote the three dimensional volume of the system, $N$ the number of constituent particles, and $m$ the mass of one particle.
The Hamiltonian of the system, $H_{V,N}$, is
\eqb
\label{eq:bf.Hvn}
 H_{V,N} \defeq
  - \dfrac{1}{2m}\,\sum_{i=1}^N \triangle_i
  + \Phi(\vec{x}_1,\cdots,\vec{x}_N) \,,
\eqe
where $\vec{x}_i$ is the spatial coordinate for $i$-th particle, and the interaction potential is
\eqb
\label{eq:bf.Phi}
 \Phi(\vec{x}_1,\cdots,\vec{x}_N) \defeq
 \sum_{j=1}^{N}\,\sum_{1\le i_1 < \cdots < i_j \le N}\,
 \phi^{(j)}(\vec{x}_{i_1},\cdots,\vec{x}_{i_j}) \,,
\eqe
where $\phi^{(j)}$ is the potential of $j$-particle interaction.
In Eq.\eref{eq:bf.Phi}, it is assumed for simplicity that the $j$-particle interaction is invariant under the permutation of spatial coordinates $\vec{x}_i$, $\phi^{(j)}(\vec{x}_1,\cdots,\vec{x}_j) = \phi^{(j)}(\vec{x}_{\tau(1)},\cdots,\vec{x}_{\tau(j)})$, where $\tau$ is the permutation.
Also, assume for simplicity that $\Phi$ vanishes for sufficiently large distribution of particles,
\eqb
\label{eq:bf.zerolevel}
 \Phi \to 0 \quad \text{as\,\, $\min\limits_{i\neq j}|\vec{x}_i-\vec{x}_j| \to \infty$} \,.
\eqe
This assumption determines the zero level of energy.

Let $\ket{\psi}$ be a normalized eigen state of $H_{V,N}$, and the boundary condition be such that the wave function $\psi \defeq \left<\vec{x}_1,\cdots,\vec{x}_N|\psi\right>$ vanishes, $\psi|_{\partial V} = 0$, at the boundary of system volume $\partial V$.
Then, the system has discrete energy eigen values,
\eqb
\label{eq:bf.Ek}
E_k(V,N) \defeq \bra{k}H_{V,N}\ket{k} \quad,\quad k=1, 2, 3, \cdots \,,
\eqe
where $\ket{k}$ is the $k$-th orthonormal eigen state.
Let the quantum number $k$ be attached in increasing order of eigen value, $E_k(V,N) \le E_{k+1}(V,N)$, where the equality repeats, $E_l(V,N) = E_{l+1}(V,N) = \cdots = E_{l+(d-1)}(V,N)$, according to the degrees of degeneracy, $d$, of degenerating states.

Let ${\mathcal H}_{V,N}$ denote the Hilbert space constructed by energy eigen states, $\ket{k}$ ($k = 1, 2, \cdots$) , and $\Omega_{V,N}(U)$ denote the number of states in ${\mathcal H}_{V,N}$ defined by
\eqb
\label{eq:bf.Omega}
 \Omega_{V,N}(U) \defeq
 \text{\em ``Number of energy eigen states satisfying $E_k(V,N) \le U$''}
 = \max\limits_{E_k \le U} k \,.
\eqe

Under the above setting and notations, the statement of theorem justifying Boltzmann formula is:
\begin{thm}[Ruelle and Tasaki]
\label{thm:rt}
For the system given above, suppose the following two conditions of interaction potential $\Phi(\vec{x}_1,\cdots,\vec{x}_N)$:
\begin{description}
\item[\,\,\,Condition~A :]
Arbitrary $j$-particle interaction, $\phi^{(j)}$, becomes negative for sufficiently large distribution of $j$ particles.
That is, there exists a constant $r_A\,(>0)$, such that
\eqb
 \phi^{(j)}(\vec{x}_{i_1},\cdots,\vec{x}_{i_j}) \le 0 \quad\text{for}\quad
 r_A \le \min\limits_{k, l = 1,\cdots,j}\left| \vec{x}_{i_k} - \vec{x}_{i_l} \right| \,.
\eqe
\item[\,\,\,Condition~B :]
The potential $\Phi$ is bounded below.
That is, there exists a constant $\phi_B\,(>0)$, such that
\eqb
 \Phi(\vec{x}_1,\cdots,\vec{x}_N) \ge -N\,\phi_B \,.
\eqe
\end{description}
Then, the following two limits exist uniquely:
\begin{description}
\item[\,\,\,Result~1 :]
The ``large system limit'' of the density of ground state energy exists,
\eqb
\label{eq:bf.thm.rt.1}
 \varepsilon_g(\rho) \defeq \lim_{l.s.l.} \dfrac{E_G(V,N)}{V} \,,
\eqe
where $E_G(V,N)$ is the eigen value of ground state defined in Eq.\eref{eq:bf.Ek}, and $\lim_{l.s.l.}$ means the large system limit defined by $V\to\infty$ with fixing $\rho \defeq N/V$ at a constant value.
This limit, $\varepsilon_g(\rho)$, is bounded below and determined uniquely.
\item[\,\,\,Result~2 :]
When $\varepsilon_g(\rho)$ does not diverge to $+\infty$, the ``thermodynamic limit'' of the logarithmic density of number of states exists,
\eqb
\label{eq:bf.thm.rt.2}
 \sigma(\varepsilon,\rho) \defeq
 \lim_{t.l.} \dfrac{\ln\Omega_{V,N}(U)}{V} \,,
\eqe
where $\lim_{t.l.}$ means the thermodynamic limit defined by $V\to\infty$ with fixing $\rho \defeq N/V$ and $\varepsilon \defeq U/V \,\ge \varepsilon_g(\rho)$ at constant values.
This limit, $\sigma(\varepsilon,\rho)$, is determined uniquely.
Furthermore, $\sigma(\varepsilon,\rho)$ is concave about its arguments $(\varepsilon,\rho)$, and monotone increasing about $\varepsilon$.
\end{description}
\end{thm}
The conditions~A and~B are the sufficient conditions for the results~1 and~2. 
The result~1 gives the lower bound to $\varepsilon$ in the result~2.
This statement of theorem follows that by Tasaki~\cite{ref:tasaki.2008}, and the same contents are found in Ruelle's book~\cite{ref:ruelle.1999}.
Proof of this theorem is lengthy and sketched in App.\ref{app:bf.proof.rt}.

Given the Ruelle-Tasaki theorem, we can expect that the uniqueness, concavity, and increasing nature of $\sigma(\varepsilon,\rho)$ given in result~2 may corresponds to those properties of thermodynamic entropy summarized in Sec.\ref{sec:td.axiom}.
In order to understand the implication of result~2 on statistical mechanics, let us discuss about the Boltzmann formula.
It is usually expressed as,
\eqb
\label{eq:bf.bf.usual}
 \widetilde{S} \defeq \ln W_{V,N}(U,\delta) \,,
\eqe
where $W_{V,N}(U,\delta)$ is the number of energy eigen states satisfying, $U-\delta V < E_k(V,N) < U+\delta V$, where $\delta \ll U/V$.
Eq.\eref{eq:bf.bf.usual} is a definition of ``statistical'' entropy, $\widetilde{S}$.
Note that, if the auxiliary parameter $\delta$ is set zero, then $W_{V,N}(U,0) = 0$ for $E_k \neq U$, or $W_{V,N}(U,0) = d$ for $E_k(V,N) = U$ with degrees of degeneracy, $d$.
That is, $\ln W_{V,N}(U,0)$ diverges to $-\infty$ for the former case, and takes a positive finite value only for the latter case with degeneracy $d \ge 2$.
Obviously, $\widetilde{S}$ in Eq.\eref{eq:bf.bf.usual} becomes ill-defined as ``entropy'' at $\delta = 0$.
Hence, the auxiliary parameter $\delta$ is necessary in definition~\eref{eq:bf.bf.usual}.
However, under the Ruelle-Tasaki theorem, the following corollary gives another expression of statistical entropy which is equivalent to Eq.\eref{eq:bf.bf.usual} and does not include the auxiliary parameter $\delta$:
\begin{coro}
\label{coro:bf}
Given the result~2 of Ruelle-Tasaki theorem, Boltzmann formula in Eq.\eref{eq:bf.bf.usual} reduces to the following form at thermodynamic limit,
\eqb
\label{eq:bf.bf}
 \widetilde{S} = \ln\Omega_{V,N}(U) \,,
\eqe
where $\Omega_{V,N}(U)$ is the number of states defined in Eq.\eref{eq:bf.Omega}.
\end{coro}
Proof of this corollary is in App.\ref{app:coro}.
Eq.\eref{eq:bf.bf} is regarded as the definition of statistical entropy.

Note that there is another theoretical evidence that $\widetilde{S}$ defined in Eq.\eref{eq:bf.bf} remains constant under reversible adiabatic process~\cite{ref:tasaki.2008}.
This evidence corresponds to a part of the entropy principle (i.e. the equality in Eq.\eref{eq:fact1.entropy}\,).
Hence, by those evidence given above so far, we find a reasonable conjecture that $\sigma(\epsilon,\rho) \,\,\, (= \lim_{t.l.}\widetilde{S}/V)$ corresponds to the density of ``thermodynamic'' entropy which satisfies various properties explained in Sec.\ref{sec:td.axiom}.
Indeed, in statistical mechanics, Eq.\eref{eq:bf.bf} is regarded as statistical expression of ``thermodynamic'' entropy, $S = \ln \Omega_{V,N}(U)$.
Thus, we can say that {\em the Ruelle-Tasaki theorem is the basic property of quantum mechanics which justifies the Boltzmann formula.}

Here it is important to make the following comment: 
The statement of Ruelle-Tasaki theorem implies that the conditions~A and~B are the sufficient conditions for the validity of Boltzmann formula. 
Thus, by Ruelle-Tasaki theorem, we can not deny a possibility that there may exist a system which violates the conditions~A and/or~B but retains the Boltzmann formula. 
However, at least in laboratory systems, it seems that such a system violating conditions~A and/or~B with retaining Boltzmann formula has not been found so far. 
It seems to be empirically probable that the conditions~A and~B hold in realistic systems~\cite{note:realsystem}.

Then, it becomes interesting to search for a necessary condition for the existence of thermal equilibrium states of quantum system under consideration. 
The following theorem is useful~\cite{ref:dobrushin.1964,ref:ruelle.1999,note:dobrushin}:
\begin{thm}[Dobrushin]
\label{thm:d}
For the quantum system given at the beginning of this subsection, consider the case satisfying following presuppositions:
\begin{description}
\item[\,\,\,Presupposition~C :]
The $j$-particle interactions for $j\neq 2$ disappear, and the total interaction potential $\Phi$ is a sum of two-particle interactions, $\Phi(\vec{x}_1,\cdots,\vec{x}_N) = \sum\limits_{1\le i < j \le N} \phi^{(2)}(\vec{x}_i,\vec{x}_j)$.
\item[\,\,\,Presupposition~D :]
Introduce a differential quantity, $D_{q_1,\cdots,q_N}$, of potential $\Phi$ defined as
\eqb
 D_{q_1,\cdots,q_N} \defeq
 \lim_{l.s.l.} \dfrac{1}{V^N}\idotsint_V {\rm d}^3x_1\,\cdots\,{\rm d}^3x_N\,
 \triangle_1^{q_1}\,\cdots\,\triangle_N^{q_N}\,
 \Phi(\vec{x}_1,\cdots,\vec{x}_N) \,,
\eqe
where $\lim_{l.s.l.}$ is the large system limit defined in Ruelle-Tasaki theorem, $q_i = 0, 1, 2, \cdots$ and $\sum_{i=1}^N q_i \neq 0$ (at least one Laplacian, $\triangle_i$, operates on $\Phi$).
Then, the presupposition~D is the requirement that this quantity satisfies the relation,
\eqb
\label{eq:bf.thm.d.1}
 \dfrac{1}{N}
 \sum_{q_1+\cdots+q_N = 1}^{\infty} \gamma^{q_1+\cdots+q_N}\,D_{q_1,\cdots,q_N}
 \,=\, \text{a finite constant independent of $N$} \,,
\eqe
where $\gamma$ is an arbitrary finite constant.
\end{description}
Under these presuppositions, if the following integral, $I_V^{(2)}$, is negative at the large system limit,
\eqb
\label{eq:bf.thm.d.IV}
 I_V^{(2)} \defeq \dfrac{1}{V^2}\iint_V {\rm d}^3x_1\,{\rm d}^3x_2\, \phi^{(2)}(\vec{x}_1,\vec{x}_2) < 0 \,,
\eqe
then the ground partition function, $\Xi_V$, of the system diverges, $\Xi_V \,\to\, \infty$, at the large system limit.
(The divergence of $\Xi_V$ denotes that no thermal equilibrium state is possible for such a system.)
\end{thm}
Proof of this theorem is in Subsec.\ref{sec:bf.proof.d}.

Here we have two comments on this theorem.
First one is on the presupposition~D, which is a technical requirement for quantum system~\cite{note:dobrushin}.
Note that the quantity $D_{q_1,\cdots,q_N}$ under the presupposition~C is essentially given by the integral,
\eqb
\label{eq:bf.thm.d.2}
 \int_V {\rm d}^3x_i \, \triangle_i^q \phi^{(2)}(\vec{x}_i,\vec{x}_j)
 = \oint_{\partial V} {\rm d}^2\tilde{x}_i \,
   \widetilde{\nabla}_i\bigl(\,\triangle_i^{q-1}\phi^{(2)}(\vec{x}_i,\vec{x}_j)\,\bigr)
\eqe
where $q=1, 2, \cdots$, and the Stokes theorem is used in the equality.
Here, $\partial V$ is the boundary of the system, ${\rm d}^2\tilde{x}_i$ is the measure on $\partial V$, and $\widetilde{\nabla}_i$ is the derivative normal to $\partial V$.
Then, we find the meanings of presupposition~D that the surface integral in right-hand side of Eq.\eref{eq:bf.thm.d.2} (i.e. the derivative of $\phi^{(2)}$ at $\partial V$) is sufficiently small so as to let the summation in left-hand side of Eq.\eref{eq:bf.thm.d.1} converge to a finite value.
Such a behavior of potential seems to be physically reasonable, at least under the requirement~\eref{eq:bf.zerolevel}.

Second comment is that, even when the system is not in thermal equilibrium state, we can consider a function $\Xi_V$ which is defined formally by the same form with ground partition function using the energy eigen values $E_k(V,N)$ and particle number $N$.
Dobrushin theorem is for such a mathematical function of the system under consideration, and says that no thermal equilibrium state is possible under the condition~\eref{eq:bf.thm.d.IV}.

\begin{figure}[h]
 \begin{center}
 \includegraphics[height=45mm]{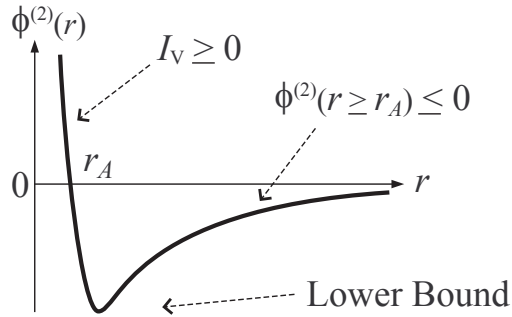}
 \end{center}
\caption{A typical potential $\phi^{(2)}(r)$ required by the existence of thermal equilibrium states.}
\label{fig:6}
\end{figure}

Finally in this subsection, let us point out the implication about interaction potential obtained from Ruelle-Tasaki and Dobrushin theorems.
To do so, the contraposition of Dobrushin theorem is useful:
\begin{description}
\item[Contraposition of Dobrushin theorem :] 
{\em Under the presuppositions~C and~D, if $\,\Xi_V$ is finite (i.e. if thermal equilibrium states exist), then $I_V^{(2)} \ge 0$ holds.}
\end{description}
The inequality $I_V^{(2)} \ge 0$ is the necessary condition for the existence of thermal equilibrium states. 
This theorem and Ruelle-Tasaki theorem imply the following:
Consider a system in which the interaction potential satisfies the sufficient conditions~A and~B of Ruelle-Tasaki theorem and the presuppositions~C and~D of Dobrushin theorem. 
By Ruelle-Tasaki theorem, the entropy of this system is well defined by the Boltzmann formula. 
This means that thermal equilibrium states of this system exist. 
Then, by (contraposition of) Dobrushin theorem together with the conditions~A and~B of Ruelle-Tasaki theorem, the two-particle interaction should be bounded below and become repulsive at a finite distance so as to let $I_V^{(2)} \ge 0$ hold.
A typical form of such a two-particle interaction is shown in Fig.\ref{fig:6}, which is very different from Newtonian gravity at short distance.
In this case, the isotropy and translational invariance of potential are assumed.
By the translational invariance, $\phi^{(2)}(\vec{x}_1,\vec{x}_2) = \phi^{(2)}(\vec{x}_1-\vec{y},\vec{x}_2-\vec{y})$ for arbitrary $\vec{y}$, which implies that $\phi^{(2)}$ is a function of only $\vec{x}_1-\vec{x}_2$, by setting $\vec{y} = \vec{x}_2$.
Then, by the isotropy, $\phi^{(2)}$ is a function of only $r = |\vec{x}_1-\vec{x}_2|$. 
Note that, the form of $\phi^{(2)}(r)$ shown in Fig.\ref{fig:6} is not the unique form but a typical form. 
The potential $\phi^{(2)}(r)$ under consideration can diverge to $+\infty$ as $r \to 0$, or can have a sufficiently large finite positive peak at smaller $r$ than that at the lower bound of $\phi^{(2)}$ so as to satisfy $I^{(2)}_V \ge 0$. 
{\em The point is that, under the conditions~A and~B together with requirement~\eref{eq:bf.zerolevel}, the potential $\phi^{(2)}(r)$ turns from attractive to repulsive as $r$ decreases around the lower bound.}

The remaining part of this section is for the proof Dobrushin theorem~\cite{note:dobrushin} which includes some original part by this author. 
A rather lengthy sketch of proof of Ruelle-Tasaki theorem is in App.\ref{app:bf.proof.rt}, which follows that of Tasaki~\cite{ref:tasaki.2008} (but slightly rearranged by this author). 
Those proofs is not necessarily needed for our conclusion, and thus readers can proceed to Sec.\ref{sec:conc} by skipping those proofs.

%%%%%%%%%%%%%%%%%%%%%%%%%%%%%%%%%%%%%%%%%%%%%%%%%%
\subsection{Proof of Dobrushin Theorem}
\label{sec:bf.proof.d}

Let us split the Hamiltonian as
\eqb
 -\beta\,H_{V,N} = \dfrac{\beta}{2 m}\,K - \beta\,\Phi \,,
\eqe
where $\beta$ is a positive constant, and $K \defeq \sum_i^N\triangle_i$.
The commutator of $K$ and $\Phi$ becomes,
\eqb
 [K,\Phi] = K\cdot\Phi - \Phi\cdot K = K[\Phi] \defeq
 \sum_{i=1}^N\triangle_i \Phi(\vec{x}_1,\cdots,\vec{x}_2) \,.
\eqe
Therefore we find,
\eqb
 [\underbrace{K,[K,\cdots,[K}_{l},\Phi]\cdots] = K^l[\Phi] \quad,\quad
 [\,\cdots\,[K,\underbrace{\Phi],\Phi],\cdots,\Phi}_{l \ge 2}] = 0 \,.
\eqe
By the Zassenhaus formula of non-commutative operators, we obtain,
\eqb
\label{eq:bf.proof.d.quantum}
 e^{-\beta H_{V,N}} =
 \exp\Bigl(\dfrac{\beta}{2m}K\Bigr)\,\exp\bigl(-\beta \Phi\bigr)\,
 \prod_{l=1}^{\infty} \exp\Bigl(\, - \beta z_l \Bigl(\dfrac{\beta}{2m}\Bigr)^l K^l[\Phi] \,\Bigr) \,,
\eqe
where $z_l$ is a numerical factor decreasing about $l$, e.g. $z_1 = -1/2$, $z_2 = 1/6$, $z_3 = -1/24\,\cdots$. 
(If we are considering a classical system, then $K$ and $\Phi$ becomes commutative, $[K,\Phi] = 0$.
This reduces the right-hand side of Eq.\eref{eq:bf.proof.d.quantum} to $e^{\beta K/2m}\,e^{\beta\Phi}$.
Then, the following part of this proof becomes more simple, and the presupposition~D is not required for classical systems.)

Then, the ground partition function $\Xi_{V,\beta,\mu}$, which is summarized in Eq.\eref{eq:prep.Xi} in App.\ref{app:prep}, becomes,
\eqab
 \Xi_{V,\beta,\mu}
 &=&
 \sum_{N=0}^{\infty} e^{\beta \mu N}\, {\rm Tr} \left[
 \exp\Bigl(\dfrac{\beta}{2m}K\Bigr)\,\exp\bigl(-\beta \Phi\bigr)\,
 \prod_{l=1}^{\infty}
 \exp\Bigl(\, - \beta z_l \Bigl(\dfrac{\beta}{2m}\Bigr)^l K^l[\Phi] \,\Bigr)\right]
\nonumber\\
 &=&
 \sum_{N=0}^{\infty} e^{\beta \mu N}
 \Bigl[\,\dfrac{1}{2\pi}\int\,{\rm d}^3p\,\exp\bigl(-\beta\dfrac{|\vec{p}|^2}{2m}\bigl)\,\Bigr]^N\,
 \idotsint_V{\rm d}^3x_1\cdots {\rm d}^3x_N \,
 e^{-\beta\Phi}\,\prod_{l=1}^{\infty}e^{-\beta z_l (\beta/2m)^l K^l[\Phi]}
\nonumber\\
 &=&
 \sum_{N=0}^{\infty}\, \Bigl( \,e^{\beta \mu}\sqrt{\dfrac{m}{2\pi\beta}}\, \Bigr)^N
  \idotsint_V{\rm d}^3x_1\cdots {\rm d}^3x_N \,
  \exp\Bigl(\, - \beta\Phi
               - \beta \sum_{l=1}^{\infty} z_l \Bigl(\dfrac{\beta}{2m}\Bigr)^l K^l[\Phi] \Bigr) \,,
\eqae
where $\mu$ is a constant, and $\vec{p}$ in the second line is the variable for momentum-representation of wave function.
Then, by the convex inequality, $(b-a)^{-1}\int_a^b {\rm d}x\,f(\,g(x)\,) \ge f\bigl(\, (b-a)^{-1}\int_a^b {\rm d}x\,g(x) \,\bigr)$, where $f(x)$ is a convex function and $g(x)$ is any arbitrary function,
\eqab
 \Xi_{V,\beta,\mu} &\ge&
 \sum_{N=0}^{\infty} \Bigl( \,e^{\beta \mu}\sqrt{\dfrac{m}{2\pi\beta}}\, \Bigr)^N\,
 V^N \exp\left[\,
 \dfrac{1}{V^N}
 \idotsint_V {\rm d}^3x_1 \cdots {\rm d}^3x_N
 \Bigl(\, - \beta \Phi
          - \beta \sum_{l=1}^{\infty} z_l \Bigl(\dfrac{\beta}{2m}\Bigr)^l K^l[\Phi] \Bigr)
 \,\right]
\nonumber \\
 &=&
 \sum_{N=0}^{\infty} \Bigl( \,e^{\beta \mu}\sqrt{\dfrac{m}{2\pi\beta}}\, \Bigr)^N\,
 V^N e^{-\beta\widetilde{D}}
 \exp\left[\,
 - \dfrac{\beta}{V^N}
 \idotsint_V {\rm d}^3x_1 \cdots {\rm d}^3x_N\, \Phi
 \,\right] \,,
\eqae
where $\widetilde{D} = \sum\limits_{q_1 + \cdots + q_N = 1}^{\infty} z_{q_1+\cdots+q_N} \Bigl(\dfrac{\beta}{2m}\Bigr)^{q_1+\cdots+q_N} D_{q_1,\cdots,q_N}$.
Note that, by the presupposition~D, there exists a constant $\omega$, such that $\widetilde{D} = \omega N$.
Furthermore, by the presupposition~C,
\eqb
 \dfrac{1}{V^N} \idotsint_V {\rm d}^3x_1 \cdots {\rm d}^3x_N\,\Phi(\vec{x}_1,\cdots,\vec{x}_N) =
 \dfrac{N(N-1)}{2 V^2}\,
 \iint_V {\rm d}^3x_1\,{\rm d}^3x_2\, \phi^{(2)}(\vec{x}_1,\vec{x}_2) \,.
\eqe
Hence, by the requirement in Eq.\eref{eq:bf.thm.d.IV}, we obtain,
\eqb
 \Xi_{V,\beta,\mu} \ge
 \sum_{N=0}^{\infty} \Bigl( \dfrac{m}{2\pi\beta} \Bigr)^{N/2}\,
 V^N
 \exp\Bigl(\,\beta\,\dfrac{N(N-1)}{2}\, \bigl| I^{(2)}_V \bigr|
             + \beta\, (\mu-\omega)\, N \,\Bigr) \,.
\eqe
The right-hand side of this inequality diverges at the large system limit, $V \to \infty$ and $N \to \infty$ with fixing $N/V$ at a constant.
Thus, the quantum version of Dobrushin theorem is proven~\cite{note:dobrushin}.
$\square$

%%%%%%%%%%%%%%%%%%%%%%%%%%%%%%%%%%%%%%%%%%%%%%%%%%%%%%%%%%%%%%%%%%%%%%%%%%%%%%%%%%%%%%%%%%%%%%%%%%%%
\section{Conclusion: Suggestion on Universal Property of Quantum Gravity}
\label{sec:conc}

We have shown the uniqueness theorem of Bekenstein-Hawking entropy (theorem~\ref{thm:uniqueness}), which is based on the entropy principle.
This means that the Bekenstein-Hawking entropy is equipped with the essential properties of entropy of ordinary laboratory systems; the entropy principle and uniqueness.
Then, it is physically reasonable to consider that the Bekenstein-Hawking entropy is calculated, in statistical mechanical sense, by applying the Boltzmann formula to a number of states determined by the underlying quantum gravity.
This may imply that the underlying quantum gravity and ordinary quantum mechanics share the same properties which justify the Boltzmann formula.
Under this consideration, we have shown the Ruelle-Tasaki theorem (theorem~\ref{thm:rt}) and quantum version of Dobrushin theorem (theorem~\ref{thm:d}).
Then, we can suggest a universal property about underlying quantum gravity as follows:

We adopt the following two basic suppositions based on black hole thermodynamics;
\begin{description}
\item[Supposition~1 :]
A stationary black hole is a stable thermal equilibrium state of microstates of underlying quantum gravity. 
(For example, the Schwarzschild black hole in the system shown in Fig.\ref{fig:2} is in a stable thermal equilibrium state whose state variables are those given in Subsec.\ref{sec:td.bh}.)
\item[Supposition~2 :]
Statistical mechanics is applicable to the thermal system of black hole.
\end{description}
From supposition~1, as shown in Sec.\ref{sec:td}, the Bekenstein-Hawking entropy possesses the essential properties of entropy; the entropy principle and uniqueness. 
From supposition~2, the Bekenstein-Hawking entropy is expressed by the Boltzmann formula which yields the entropy {\em uniquely}. 
Concerning the Boltzmann formula, recall that the conditions~A and~B of Ruelle-Tasaki theorem are the sufficient conditions for the validity of Boltzmann formula at least for laboratory systems. 
Furthermore, note that there seems to be no example, at least for laboratory systems, which does not satisfy the conditions~A and/or~B but retains the Boltzmann formula. 
Thus, it seems to be empirically reasonable to consider that the conditions~A and~B of Ruelle-Tasaki theorem holds also in the underlying quantum gravity. 
If it is true, then the possible suggestion is:
\begin{description}
\item[Suggestion~1 :]
The interaction potential among microstates of underlying quantum gravity is bounded below, unlike the Newtonian gravity. 
This lower bound is given at Planck length scale, because the general relativity or Newtonian approximation of gravity should be recovered for length scale larger than Planck size.
\end{description}
On the other hand, by the existence of black hole which is thermal equilibrium state of gravity as mentioned in supposition~1, it is reasonable to consider that Dobrushin theorem holds also in quantum gravity. 
Then, if the suggestion~1 holds, the necessary condition for the existence of thermal equilibrium state, which is shown in Dobrushin theorem, gives the following suggestion:
\begin{description}
\item[Suggestion~2 :]
If the interaction of underlying quantum gravity is a sum of two-body interaction, then the two-body interaction becomes repulsive at Planck length scale, as shown in Fig.\ref{fig:6}.
Quantum gravity may become repulsive at Planck length.
\end{description}
Here, there arises an issue about what the ``interaction potential'' means in quantum gravity, since it is not necessarily clear whether the quantum gravitational interaction is to be expressed by the interaction potential such as $\phi^{(2)}(r)$ in Hamiltonian~\eref{eq:bf.Hvn}. 
In the case that the full quantum effect of gravity is hard to be expressed by the interaction potential, the above suggestions can be understood as follows: 
\begin{description}
\item[An interpretation of suggestions :]
When the full quantum gravity is approximated to a ``semi-classical'' gravity, the semi-classical correction to Einstein-Hilbert action may be restricted so as to cause the repulsive gravity around Planck length scale.
\end{description} 
Finally, let us emphasize that the above suggestions and interpretation about underlying quantum gravity are universal in the sense that they are independent of any existing model of quantum gravity (e.g. superstring theory, loop quantum gravity, (causal) dynamical triangulation, and so on), since all proofs of theorems referred to in this paper do not use any exiting model of quantum gravity.

%%%%%%%%%%%%%%%%%%%%%%%%%%%%%%%%%%%%%%%%%%%%%%%%%%%%%%%%%%%%%%%%%%%%%%%%%%%%%%%%%%%%%%%%%%%%%%%%%%%%
\section*{Acknowledgements}

This work is supported by the Grant-in-Aid for Scientific Research Fund of the Ministry of Education, Culture, Sports, Science and Technology, Japan [Young Scientists (B) 19740149], and also by the grant of Daiko Foundation [No.9130].

%%%%%%%%%%%%%%%%%%%%%%%%%%%%%%%%%%%%%%%%%%%%%%%%%%%%%%%%%%%%%%%%%%%%%%%%%%%%%%%%%%%%%%%%%%%%%%%%%%%%
\appendix
%%%%%%%%%%%%%%%%%%%%%%%%%%%%%%%%%%%%%%%%%%%%%%%%%%%%%%%%%%%%%%%%%%%%%%%%%%%%%%%%%%%%%%%%%%%%%%%%%%%%
\section{Construction of Free Energy of Schwarzschild Black Hole}
\label{app:free}

This appendix is for the construction of free energy~\eref{eq:td.Fbh} without using the Euclidean quantum gravity.
Instead of referring to Euclidean quantum gravity, we refer to a general relativistic (classical) property of black hole and the Hawking radiation.
Consider the case that single Schwarzschild black hole is in an empty spacetime, which is exactly described by the line element in Eq.\eref{eq:td.metric}.
Then, by evaluating the Noether charge of black hole spacetime at spatial infinity and at horizon, the following differential relation is obtained~\cite{ref:bardeen+2.1973,ref:iyer+1.1994},
\eqb
\label{eq:free.1st}
 {\rm d}M = \dfrac{\kappa}{2\pi}\,{\rm d}(4 \pi M^2) \,,
\eqe
where $\kappa = 1/(4M)$ is the surface gravity of black hole horizon.
The left hand-side, ${\rm d}M$, comes from the Noether charge at spatial infinity.
The differential quantity, ${\rm d}(4 \pi M^2)$, comes from the Noether charge at black hole horizon.
Eq.\eref{eq:free.1st} is purely a geometrical relation.

Then, introduce a quantum field on the Schwarzschild spacetime.
We find that the black hole emits the Hawking radiation with thermal spectrum~\cite{ref:hawking.1975}.
Based on this thermal spectrum, we adopt the idea that {\em the black hole is a thermal equilibrium state of gravitational field}.
Furthermore, for an infinitely distant observer from black hole, the temperature of Hawking radiation is $(8\pi M)^{-1}$, which coincides with two quantities; the factor in right-hand side of Eq.\eref{eq:free.1st}, and the limit of $T_{\rm BH}$ in Eq.\eref{eq:td.Tbh}, $T_{\rm BH} \to (8\pi M)^{-1}$ as $r_w \to \infty$.
By this coincidence, we adopt the idea that, for the thermal system of black hole shown in Fig.\ref{fig:2}, Eq.\eref{eq:free.1st} is the limit form of first law of black hole thermodynamics as $r_w\to\infty$.
Thus, it may be reasonable to require the followings for the thermal system of black hole shown in Fig.\ref{fig:2}:
\begin{description}
\item[Requirement~1:]
The internal energy of black hole, $U_{\rm BH}$, becomes $M$ at the limit of distant observer, $\lim\limits_{r_w \to \infty}U_{\rm BH} = M$, since the term ${\rm d}M$ in Eq.\eref{eq:free.1st} comes from the Noether charge at spatial infinity.
\item[Requirement~2:]
The entropy of black hole is $4 \pi M^2$ for arbitrary $r_w$, since the factor ${\rm d}(4\pi M^2)$ comes from the Noether charge at black hole horizon (independent of $r_w$ ).
\end{description}
We construct the free energy using these requirements together with $A_w$ in Eq.\eref{eq:td.Aw} and $T_{\rm BH}$ in Eq.\eref{eq:td.Tbh}.

By the requirement~2 and ordinary thermodynamic relation, the desired free energy $F_{\rm BH}$ should satisfy,
\eqb
\label{eq:free.entropy}
 -\,\pd{F_{\rm BH}(T_{\rm BH},A_w)}{T_{\rm BH}} \,=\, 4 \pi M^2 \,,
\eqe
where following ordinary thermodynamics, $F_{\rm BH}$ is regarded as a function of temperature and system size.
By definition of $A_w$, the left-hand side of Eq.\eref{eq:free.entropy} is expressed by the partial derivatives of $F_{\rm BH}$ and $T_{\rm BH}$ about $M$.
Then, integrating Eq.\eref{eq:free.entropy} by $M$,
\eqb
 F_{\rm BH}
 = \int {\rm d}M (-4\pi M^2)\,\pd{T_{\rm BH}(M,r_w)}{M}
 = - r_w\,\sqrt{1-\dfrac{2M}{r_w}} - \dfrac{M}{2\sqrt{1-2M/r_w}} + f(r_w) \,,
\eqe
where $f(r_w)$ is arbitrary function of $r_w$.

Following ordinary thermodynamics, the free energy $F_{\rm BH}$ and internal energy $U_{\rm BH}$ are related by the Legendre transformation, $U_{\rm BH}(S_{\rm BH},A_w) = F_{\rm BH}(T_{\rm BH},A_w) + T_{\rm BH}\,S_{\rm BH}$, where $S_{\rm BH} = 4\pi M^2$ is the entropy and $T_{\rm BH}$ is regarded as a function of $S_{\rm BH}$ and $A_w$.
This denotes that thermodynamic energies, such as $U_{\rm BH}$ and $F_{\rm BH}$, have the same scaling behavior with $T_{\rm BH}\,S_{\rm BH}$ under the scaling of fundamental parameters, $M \to \lambda M$ and $r_w \to \lambda r_w$, where $\lambda\,(> 0)$ is the scaling rate of length size of system.
($M$ and $r_w$ have the dimension of length.)
Because of the scaling behavior, $T_{\rm BH}\,S_{\rm BH} \to \lambda T_{\rm BH}\,S_{\rm BH}$, under those fundamental scalings, the free energy should be scaled as $F_{\rm BH} \to \lambda\,F_{\rm BH}$.
This implies $f(r_w) = a\,r_w$, where $a$ is a constant.

From the above, we find $U_{\rm BH} = F_{\rm BH} + T_{\rm BH}\,S_{\rm BH} = r_w\,(a-\sqrt{1-2M/r_w}\,)$, which becomes,
\eqb
 U_{\rm BH} \to (a-1)\,r_w + M\,\Bigl[\,1 + O\Bigl(\dfrac{M}{r_w}\Bigr)\,\Bigr]
 \quad \text{as $r_w\to\infty$.}
\eqe
Then, by the requirement~1, we obtain $a = 1$ and $F_{\rm BH}$ becomes the form given in Eq.\eref{eq:td.Fbh}.

Here it is helpful to point out that the requirement 2 can also be regarded as the {\em conceptual basis} of the Euclidean quantum gravity in the following sense: 
In the Euclidean quantum gravity, the Euclidean action integral, which is obtained from the Lorentzian action via the so-called Wick rotation, is regarded as the partition function of the thermal system under consideration. 
Although the calculation in Euclidean quantum gravity can be carried out under the basic assumption that the Euclidean action corresponds to the partition function, however the validity of the calculation (i.e. the validity of the basic assumption) can not be checked in, solely, the framework of Euclidean quantum gravity. 
It is the requirement~2 that has been originally referred to in order to infer the validity of Euclidean quantum gravity~\cite{ref:gibbons+1.1977}. 
That is, because the Euclidean quantum gravity reproduces the black hole entropy $S_{\rm BH} = 4\pi M^2$ which is the theoretically reasonable form of entropy based on general relativity and quantum field theory in curved spacetime, we can accept the Euclidean quantum gravity as one candidate of possible theories of underlying quantum gravity. 
In this sense, the requirement~2 is the {\em conceptual basis} of the Euclidean quantum gravity, while the theoretical basis of Euclidean quantum gravity is to regard the Euclidean action as the partition function.

In this paper, instead of regarding the Euclidean action as the partition function (i.e. instead of using the Euclidean quantum gravity), we adopt not only the requirement~2 but also the requirement~1. 
Then, as shown above, the free energy in Eq.\eref{eq:td.Fbh} is obtained. 
This means that {\em the black hole thermodynamics can be established by not only the Euclidean quantum gravity but also any model of quantum gravity satisfying the requirements~1 and~2.} 
In this sense, the discussion in this paper does not depend on details of existing models of quantum gravity, but simply requires the requirements~1 and~2 at least for Schwarzschild black hole.

Finally in this appendix, recall that the requirements~1 and~2 describe some properties of state variables of thermal system of black hole. 
Hence, rigorously speaking, those requirements imply also the {\em existence} of thermal equilibrium state of black hole (such as the system in Fig.\ref{fig:2}), because state variables can not be defined unless thermal equilibrium is possible. 
In the conclusion of this paper in Sec.\ref{sec:conc}, the requirements~1 and~2 are included into the "supposition~1" which requires the existence of thermal equilibrium system of black hole.

%%%%%%%%%%%%%%%%%%%%%%%%%%%%%%%%%%%%%%%%%%%%%%%%%%
\section{Sketch of Proof of Ruelle-Tasaki Theorem}
\label{app:bf.proof.rt}

This appendix follows basically the appendix~C in Tasaki's book~\cite{ref:tasaki.2008}. 
The substep~3-2 in the following proof is constructed by this author, since it is not explicitely given in Tasaki's book but left as an ``exercise'' for readers. 
To show the substep~3-2, the detail of explanation in this paper is not exactly the same with that in Tasaki's book, but slightly rearranged by this author.

%%%%%%%%%%
\subsection{Preparations}

Introduce two propositions used in the proof of Ruelle-Tasaki theorem. 
For the first, let us show a mathematical fact about eigen values:

\begin{prop}[Mini-max principle]
\label{prop:minimax}
Consider the quantum system supposed in the statement of Ruelle-Tasaki theorem, in which the ground state energy is bounded below ($E_G(V,N) =$ finite) due to the condition~B.
Let ${\mathcal C}_n$ be an n-dimensional subspace in Hilbert space ${\mathcal H}_{V,N}$ of the system, and $\Theta_H[{\mathcal C}_n]$ denote the supremum of energy expectation value, $\bra{\psi_n} H_{V,N} \ket{\psi_n}$, of arbitrary normalized state $\ket{\psi_n}$ in ${\mathcal C}_n$,
\eqb
\label{eq:bf.minimax.Theta}
 \Theta_H[{\mathcal C}_n] \defeq
 \sup_{\substack{\ket{\psi_n} \,\in\, {\mathcal C}_n \\
                 \bracket{\psi_n} = 1}}
 \bra{\psi_n} H_{V,N} \ket{\psi_n} \,.
\eqe
Then, the $k$-th energy eigen value $E_k(V,N)$ is equal to the infimum of $\Theta_H[{\mathcal C}_k]$ such as,
\eqb
 E_k(V,N) = \inf_{{\mathcal C}_k} \Theta_H[{\mathcal C}_k] \,,
\eqe
where the infimum is evaluated under the variation of ${\mathcal C}_k$ in ${\mathcal H}_{V,N}$ with fixing its dimension at $k$.
\end{prop}
Proof of this proposition is found in textbooks of functional analysis and mathematical foundation of quantum mechanics, or see Ruelle's book~\cite{ref:ruelle.1999} for example.

Next, we show a proposition about the number of states:
\begin{prop}
\label{prop:interacting}
For arbitrary constant $\tilde{\beta}\,(>0)$, the number of states $\Omega_{V,N}(U)$ of our quantum system is bounded above at the large system limit (defined in the statement of Ruelle-Tasaki theorem, result~1),
\eqb
 \Omega_{V,N}(U) \le
 \exp\bigl(\, \tilde{\sigma}\,V + \tilde{\beta}\,U \,\bigr) \,,
\eqe
where $\tilde{\sigma}(\rho,\tilde{\beta})$ is a constant determined by $\tilde{\beta}$ and $\rho \defeq N/V$ which is the number density fixed in the large system limit.
\end{prop}
Proof of this proposition is in App.\ref{app:interacting} in which two lemmas, which are summarized in App.\ref{app:prep}, are used.

These two propositions are used in the proof of Ruelle-Tasaki theorem.
However, the proof shown below is not necessarily mathematically rigorous.
(The mathematical detail of proof is found in Ruelle's book~\cite{ref:ruelle.1999}.)
Let us show the central idea of the proof following Tasaki's book~\cite{ref:tasaki.2008}.
The idea of proof is divided into some steps as follows:
\begin{description}
\item[\,\,\,Step~1 :]
Introduce a basic technique used in the proof. (Mini-max principle is used.)
\item[\,\,\,Step~2 :]
Show the result~1 of Ruelle-Tasaki theorem.
\item[\,\,\,Step~3 :]
Show the result~2 of Ruelle-Tasaki theorem.
This step consists of three substeps:
\begin{description}
\item[\rm Substep 3-1 :]
Show the existence of unique thermodynamic limit, $\sigma(\varepsilon,\rho)$. (Proposition~\ref{prop:interacting} is used.)
\item[\rm Substep 3-2 :]
Show the concavity of $\sigma(\varepsilon,\rho)$ about $\varepsilon$ and $\rho$.
\item[\rm Substep 3-3 :]
Show the monotone increasing nature of $\sigma(\varepsilon,\rho)$ about $\varepsilon$.
\end{description}
\end{description}

%%%%%%%%%%
\subsection{Step~1 of the Proof: basic technique}

Consider two systems of identical particles in spatial regions, ${\mathcal D}^{(a)}$ and ${\mathcal D}^{(b)}$, and the number of particles are $N^{(a)}$ and $N^{(b)}$, respectively, in ${\mathcal D}^{(a)}$ and ${\mathcal D}^{(b)}$.
These systems satisfy the conditions~A and~B.
Let the distance, $L$, between ${\mathcal D}^{(a)}$ and ${\mathcal D}^{(b)}$ satisfy $L \ge r_A$, where $r_A$ is given in the condition~A.
Use the same notations for various quantities of these systems as given in the beginning of Sec.\ref{sec:bf.thm}, for example the eigen value $E_k^{(a)}(V^{(a)},N^{(a)})$, number of states $\Omega^{(b)}_{V^{(b)},N^{(b)}}(U^{(b)})$, and so on.
Here, the Hamiltonian of each system is
\eqb
\label{eq:bf.proof.rt.1.Hi}
 H^{(i)}_{V^{(i)},N^{(i)}} \defeq
 - \dfrac{1}{2 m}\sum_{l=1}^{N^{(i)}}\triangle_l
 + \Phi^{(i)}(\vec{x}_1^{(i)},\cdots,\vec{x}_{N^{(i)}}^{(i)}) \,,
\eqe
where $\vec{x}_k^{(i)} \in {\mathcal D}^{(i)}$, and $i = a, b$.
For the later use, define a subspace ${\mathcal C}^{(i)}[U^{(i)}] \,\,\, (i = a, b)$ in each Hilbert space ${\mathcal H}_{V^{(i)},N^{(i)}}^{(i)}$ of these systems by
\eqb
\label{eq:bf.proof.rt.1.Ci}
 {\mathcal C}^{(i)}[U^{(i)}] \defeq
 \left\{\,
  \text{linear combinations of energy eigen states $\ket{k}^{(i)}$} \,\,\Bigr|\,\,
  k = 1,\cdots,\Omega_{V^{(i)},N^{(i)}}^{(i)}(U^{(i)})
 \,\right\} \,.
\eqe
Obviously, its dimension is the number of energy eigen values lower than or equal to $U^{(i)}$, $\dim {\mathcal C}^{(i)}[U^{(i)}] = \Omega_{V^{(i)},N^{(i)}}^{(i)}(U^{(i)})$.
For any arbitrary normalized state in ${\mathcal C}^{(i)}[U^{(i)}]$ given as,
\eqb
\label{eq:bf.proof.rt.1.psii}
 \ket{\psi^{(i)}[U^{(i)}]}
 =
 \sum_{k=1}^{\dim {\mathcal C}^{(i)}[U^{(i)}]}\,
 \alpha_k\,\ket{k}^{(i)} \qquad,\qquad
 \sum_k |\alpha_k|^2 = 1 \,,
\eqe
we find an inequality,
\eqb
\label{eq:bf.proof.rt.1.ineq.i}
 \bra{\psi^{(i)}[U^{(i)}]}H^{(i)}_{V^{(i)},N^{(i)}}\ket{\psi^{(i)}[U^{(i)}]}
 = \sum_{k=1}^{\dim {\mathcal C}^{(i)}[U^{(i)}]} |\alpha_k|^2\,E_k^{(i)}(V^{(i)},N^{(i)})
 \le U^{(i)} \,.
\eqe
Hence, ${\mathcal C}^{(i)}[U^{(i)}]$ is composed of states which have energy lower than or equal to $U^{(i)}$.

When we regard these two systems as one total system such as the composition of macroscopic systems (see definition~\ref{def:composition} in Sec.\ref{sec:td.axiom}), the total Hamiltonian is
\eqb
\label{eq:bf.proof.rt.1.Htot}
 H^{\rm tot}_{V^{\rm tot},N^{\rm tot}} =
 H^{(a)}_{V^{(a)},N^{(a)}} + H^{(b)}_{V^{(b)},N^{(b)}} + \Phi_{\rm int} \,,
\eqe
where $V^{\rm tot} = V^{(a)} + V^{(b)}$, $N^{\rm tot} = N^{(a)} + N^{(b)}$, and $\Phi_{\rm int}$ is the interaction between the two subsystems,
\eqb
\label{eq:bf.proof.rt.1.Phiint}
 \Phi_{\rm int} \defeq
 \sum_{n=2}^{N^{\rm tot}}\,
 \sum_{1\le i_1 < \cdots < i_n \le N^{\rm tot}}\,
 \phi^{(n)}(\vec{x}_{i_1},\cdots,\vec{x}_{i_n})\, \chi_{i_1,\cdots,i_n} \,,
\eqe
where
\eqb
\label{eq:bf.proof.rt.1.chi}
 \chi_{i_1,\cdots,i_n} =
 \begin{cases}
  0 &\text{:\, All $\vec{x}_l$ ($l = i_1,\cdots,i_n$) are in the same subsystem}\\
  1 &\text{:\, The other cases}
 \end{cases} \,.
\eqe
By the requirement $L \ge r_A$ and the condition~A, we have $\Phi_{\rm int} \le 0$.
Let $E^{\rm tot}_k(V^{\rm tot},N^{\rm tot})$ denote the $k$-th energy eigen value of the total Hamiltonian~\eref{eq:bf.proof.rt.1.Htot}.

Define a subspace ${\mathcal C}^{\rm tot}\bigl[\widetilde{U}\bigr]$ in the total Hilbert space ${\mathcal H}_{V^{\rm tot},N^{\rm tot}}^{\rm tot}$ by,
\eqb
\label{eq:bf.proof.rt.1.Ctot}
 {\mathcal C}^{\rm tot}\bigl[\widetilde{U}\bigr] \defeq
 {\mathcal C}^{(a)}[U^{(a)}] \,\widehat{\otimes}\, {\mathcal C}^{(b)}[U^{(b)}] \,,
\eqe
where $\widetilde{U} = U^{(a)} + U^{(b)}$, and $\widehat{\otimes}$ is the anti-symmetrized product if subsystems are fermionic or the symmetrized product if subsystems are bosonic.
Obviously, its dimension is
\eqb
\label{eq:bf.proof.rt.1.dimCtot}
 \dim {\mathcal C}^{\rm tot}\bigl[\widetilde{U}\bigr] =
 \dim {\mathcal C}^{(a)}[U^{(a)}] \cdot \dim {\mathcal C}^{(b)}[U^{(b)}] =
 \Omega_{V^{(a)},N^{(a)}}^{(a)}(U^{(a)}) \cdot \Omega_{V^{(b)},N^{(b)}}^{(b)}(U^{(b)}) \,.
\eqe
Let $\ket{\psi^{\rm tot}\bigl[\widetilde{U}\bigr]}$ be any arbitrary normalized state in ${\mathcal C}^{\rm tot}\bigl[\widetilde{U}\bigr]$, which, by definition~\eref{eq:bf.proof.rt.1.Ctot}, is given by
\eqb
\label{eq:bf.proof.rt.1.psitot}
 \ket{\psi^{\rm tot}\bigl[\widetilde{U}\bigr]} =
 \ket{\psi^{(a)}[U^{(a)}]} \widehat{\otimes} \ket{\psi^{(b)}[U^{(b)}]} \,.
\eqe
Then, from $\Phi_{\rm int} \le 0$ and Eq.\eref{eq:bf.proof.rt.1.ineq.i}, we obtain,
\eqb
\label{eq:bf.proof.rt.1.ineq.tot}
\begin{split}
 \bra{\psi^{\rm tot}\bigl[\widetilde{U}\bigr]} & H_{V^{\rm tot},N^{\rm tot}}^{\rm tot}
 \ket{\psi^{\rm tot}\bigl[\widetilde{U}\bigr]} \\
 &=
 \bra{\psi^{(a)}[U^{(a)}]}H^{(a)}_{V^{(a)},N^{(a)}}\ket{\psi^{(a)}[U^{(a)}]}
  + \bra{\psi^{(b)}[U^{(b)}]}H^{(b)}_{V^{(b)},N^{(b)}}\ket{\psi^{(b)}[U^{(b)}]}
  + \Phi_{\rm int} \\
 &\le U^{(a)} + U^{(b)} \,.
\end{split}
\eqe
This inequality implies that the supremum quantity defined in Eq.\eref{eq:bf.minimax.Theta} is bounded above, $\Theta_{H^{\rm tot}}[C^{\rm tot}] \le \widetilde{U}$.
Therefore, by the mini-max principle (proposition~\ref{prop:minimax}), we obtain an upper bound of energy eigen value,
\eqb
\label{eq:bf.proof.rt.1.Etot}
 E^{\rm tot}_{\dim {\mathcal C}^{\rm tot}[U^{(a)} + U^{(b)}]}(V^{\rm tot},N^{\rm tot})
 \le U^{(a)} + U^{(b)} \,.
\eqe
This inequality implies,
\eqb
 \dim {\mathcal C}^{\rm tot}[U^{(a)} + U^{(b)}]
 \le \Omega_{V^{\rm tot},N^{\rm tot}}^{\rm tot}(U^{(a)} + U^{(b)}) \,,
\eqe
where $\Omega_{V^{\rm tot},N^{\rm tot}}^{\rm tot}(U^{(a)} + U^{(b)})$ is the number of states in total system.
Hence, by Eq.\eref{eq:bf.proof.rt.1.dimCtot}, we obtain
\eqb
\label{eq:bf.proof.rt.1.Omegatot}
 \Omega_{V^{(a)},N^{(a)}}^{(a)}(U^{(a)}) \cdot \Omega_{V^{(b)},N^{(b)}}^{(b)}(U^{(b)})
 \le \Omega_{V^{\rm tot},N^{\rm tot}}^{\rm tot}(U^{(a)} + U^{(b)}) \,.
\eqe
The inequalities of energy eigen value~\eref{eq:bf.proof.rt.1.Etot} and of number of states~\eref{eq:bf.proof.rt.1.Omegatot} are used in following steps.

%%%%%%%%%%
\subsection{Step~2 of the Proof: result~1}

\begin{figure}[h]
 \begin{center}
 \includegraphics[height=65mm]{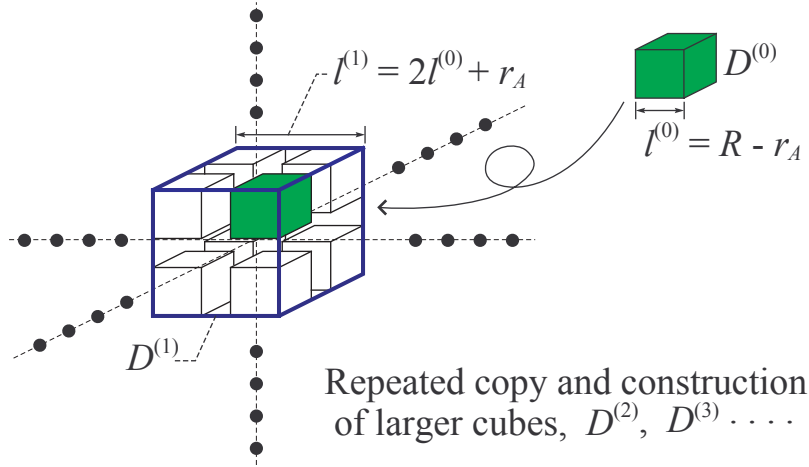}
 \end{center}
\caption{Construction of a large system.}
\label{fig:7}
\end{figure}

Let us construct a large system by the following procedure:
\begin{itemize}
\item[(i)]
Let ${\mathcal D}^{(0)}$ be a cubic region of edge length $l^{(0)} \defeq R-r_A$, where $R\,(>r_A)$ is a constant, in which $N^{(0)}$ identical particles exist.
Let $V^{(0)}$ denote the volume of this cube, $V^{(0)} = l^{(0)\,3} = (R-r_A)^3$.
The Hamiltonian of this system, $H^{(0)}_{V^{(0)},N^{(0)}}$, is expressed as that in Eq.\eref{eq:bf.proof.rt.1.Hi}.
Require that the conditions~A and~B are satisfied.
\item[(ii)]
Let ${\mathcal D}^{(1)}$ be a cubic region of edge length $l^{(1)} \defeq 2 l^{(0)} + r_A = 2R - r_A$, and $V^{(1)}$ denote its volume, $V^{(1)} = l^{(1)\, 3}$.
Then, make eight copies of the cube ${\mathcal D}^{(0)}$ (including $N^{(0)}$ particles), and place them inside ${\mathcal D}^{(1)}$ as shown in Fig.\ref{fig:7} so as to share the eight vertices of ${\mathcal D}^{(1)}$ with the eight copies of ${\mathcal D}^{(0)}$.
By this construction of larger cube ${\mathcal D}^{(1)}$, the distance between smaller cubes ${\mathcal D}^{(0)}$ is longer than or equal to $r_A$.
In the larger cube ${\mathcal D}^{(1)}$, there exist $8 N^{(0)}$ particles.
The Hamiltonian of this system, $H^{(1)}_{V^{(1)},N^{(1)}}$, is expressed as that in Eq.\eref{eq:bf.proof.rt.1.Htot},
\eqb
 H^{(1)}_{V^{(1)},N^{(1)}} = 8 H^{(0)}_{V^{(0)},N^{(0)}} + \Phi_{\rm int}^{(1)} \,,
\eqe
where $\Phi_{\rm int}^{(1)}$ is defined as that in Eq.\eref{eq:bf.proof.rt.1.Phiint}.
By the condition~A, $\Phi_{\rm int}^{(1)} \le 0$ holds.
\item[(iii)]
Let ${\mathcal D}^{(2)}$ be a cubic region of edge length $l^{(2)} \defeq 2 l^{(1)} + r_A = 4R - r_A$, and $V^{(2)}$ denote its volume.
Repeat the procedure~(ii) and construct the larger system in ${\mathcal D}^{(2)}$ including $8^2 N^{(0)}$ particles with Hamiltonian, $H^{(2)}_{V^{(2)},N^{(2)}} = 8 H^{(1)}_{V^{(1)},N^{(1)}} + \Phi_{\rm int}^{(2)}$, where $\Phi_{\rm int}^{(2)} \le 0$.
Then, repeating again the same procedure $n$ times, the $n$-th cube ${\mathcal D}^{(n)}$ of edge length $l^{(n)} = 2^n R - r_A$ is constructed, which includes $8^n N^{(0)}$ particles with Hamiltonian, $H^{(n)}_{V^{(n)},N^{(n)}} = 8^n H^{(0)}_{V^{(0)},N^{(0)}} + \sum_{i=1}^n 8^i \Phi_{\rm int}^{(i)}$, where $\Phi_{\rm int}^{(i)} \le 0$.
For sufficiently large $n$, we obtain a large system.
Obviously, the inequalities~\eref{eq:bf.proof.rt.1.Etot} and~\eref{eq:bf.proof.rt.1.Omegatot} can be applied to this large system with appropriate modifications.
\end{itemize}

In the large system constructed by the above procedure, consider an $n$-th cube ${\mathcal D}^{(n)}$ which includes eight $(n-1)$-th cubes ${\mathcal D}^{(n-1)}$.
By repeating the same calculation to obtain inequality~\eref{eq:bf.proof.rt.1.Etot}, we obtain
\eqb
\label{eq:bf.proof.rt.2.En}
 E^{(n)}_{\dim {\mathcal C}^{(n)}[8 U^{(n-1)}]}(V^{(n)},N^{(n)})
 \le 8 U^{(n-1)} \,,
\eqe
where ${\mathcal C}^{(n)}[8 U^{(n-1)}] = {\mathcal C}^{(n-1)}[U^{(n-1)}]\widehat{\otimes} \cdots \widehat{\otimes}{\mathcal C}^{(n-1)}[U^{(n-1)}]$ (eight products) is a subspace in $n$-th Hilbert space ${\mathcal H}^{(n)}_{V^{(n)},N^{(n)}}$.

Consider the case that $U^{(n-1)}$ is the ground state energy, $E_G^{(n-1)}(V^{(n-1)},N^{(n-1)})$, of the system in an $(n-1)$-th cube ${\mathcal D}^{(n-1)}$.
By definition~\eref{eq:bf.proof.rt.1.Ci}, the subspace ${\mathcal C}^{(n-1)}[E_G^{(n-1)}]$ in ${\mathcal H}^{(n-1)}_{V^{(n-1)},N^{(n-1)}}$ is spanned by the ground states, $\ket{1}^{(n-1)},\cdots,\ket{d}^{(n-1)}$, where $d$ is the degrees of degeneracy at ground state.
Then, as implied by Eq.\eref{eq:bf.proof.rt.1.Ctot}, the subspace ${\mathcal C}^{(n)}[8E_G^{(n-1)}]$ is spanned by the states, $\ket{(k_1,\cdots,k_8)}^{(n)} \defeq \widehat{\otimes}_{i=1}^8\ket{k_i}^{(n-1)}$, $(k_i = 1,\cdots,d)$.
This state, $\ket{(k_1,\cdots,k_8)}^{(n)}$, is a ground state in ${\mathcal H}^{(n)}_{V^{(n)},N^{(n)}}$.
Hence, the left-hand side in Eq.\eref{eq:bf.proof.rt.2.En} becomes the ground state energy, $E_G^{(n)}(V^{(n)},N^{(n)})$, of the system in $n$-th cube ${\mathcal D}^{(n)}$.
Rearranging Eq.\eref{eq:bf.proof.rt.2.En}, we obtain
\eqb
\label{eq:bf.proof.rt.2.EGn}
 \dfrac{E_G^{(n)}(V^{(n)},N^{(n)})}{(2^n R)^3}
 \le
 \dfrac{E_G^{(n-1)}(V^{(n-1)},N^{(n-1)})}{(2^{n-1} R)^3} \,.
\eqe
This denotes that the sequence, $f_n \defeq (2^n R)^{-3} E_G^{(n)}(V^{(n)},N^{(n)})$, is decreasing about $n$, $f_{n-1} \ge f_n$.
Therefore, $f_n$ diverges to $-\infty$ or converges to a {\em unique} constant, as $n \to \infty$.
On one hand, $f_n$ should be bounded below due to the condition~B,
\eqb
 \dfrac{E_G^{(n)}(V^{(n)},N^{(n)})}{(2^n R)^3} =
 \dfrac{1}{(2^n R)^3}
 \,^{(n)}\bra{(k_1,\cdots,k_8)} H^{(n)}_{V^{(n)},N^{(n)}} \ket{(k_1,\cdots,k_8)}^{(n)}
 \ge - \dfrac{N^{(n)}}{(2^n R)^3}\,\phi_B \,,
\eqe
where $N^{(n)}/(2^n R)^3 = N^{(0)}/R^3$ is a constant.
Hence, there exists a unique limit, $f_{\infty} = \lim_{n\to\infty}f_n$.
In the above discussion, there remains a possibility that $f_{\infty} = +\infty$.

Furthermore, by definition $V^{(n)} \defeq l^{(n)\,3} = (2^n R -r_A)^3$, we find $(2^n R)^3/V^{(n)} \to 1$ as $n\to\infty$, which means that the density of ground state energy, $E^{(n)}_G(V^{(n)},N^{(n)})/V^{(n)} = f_n (2^n R)^3/V^{(n)}$, has a unique limit as $n\to\infty$.
Also, by definition $N^{n} \defeq 8^n N^{(0)}$, we find $N^{(n)}/V^{(n)} \to N^{(0)}/R^3$ as $n\to\infty$, which means that the limit operation, $n\to\infty$, of the large system considered here is the ``large system limit'' required in the statement of theorem.
Thus, it is proven that there exists a unique large system limit of the density of ground state energy, $\varepsilon_g(\rho)$, as expressed in Eq.\eref{eq:bf.thm.rt.1}, where $\rho \defeq N^{(0)}/V^{(0)}$.
The result~1 is (roughly) proven~\cite{note:proof.rt}.

%%%%%%%%%%
\subsection{Step~3 of the Proof: result~2}

%%%%%%%%%%
\subsubsection{Substep~3-1}

Consider the same large system with the step~2.
Then, for an $n$-th cube ${\mathcal D}^{(n)}$ which is composed of eight $(n-1)$-th cubes ${\mathcal D}^{(n-1)}$, we obtain, by repeating the same calculation to obtain inequality~\eref{eq:bf.proof.rt.1.Omegatot},
\eqb
 \Bigl[\, \Omega^{(n-1)}_{V^{(n-1)},N^{(n-1)}}\bigl(U^{(n-1)}\bigr) \,\Bigr]^8
 \le
 \Omega^{(n)}_{V^{(n)},N^{(n)}}\bigl(8U^{(n-1)}\bigr) \,.
\eqe
Take the logarithm and divide it by $(2^n R)^3$,
\eqb
\label{eq:bf.proof.rt.3.Omegan}
 \dfrac{\ln \Omega^{(n-1)}_{V^{(n-1)},N^{(n-1)}}\bigl(U^{(n-1)}\bigr)}{(2^{n-1} R)^3}
 \le
 \dfrac{\ln \Omega^{(n)}_{V^{(n)},N^{(n)}}\bigl(U^{(n)}\bigr)}{(2^n R)^3} \,,
\eqe
where $U^{(n)} \defeq 8 U^{(n-1)}$, that is $U^{(n)} = 8^n U^{(0)}$.
Here, $U^{(0)}$ is the energy of system in a smallest cube ${\mathcal D}^{(0)}$, for which an inequality, $U^{(0)}\ge E^{(0)}_G(V^{(0)},N^{(0)})$, should hold.

The inequality~\eref{eq:bf.proof.rt.3.Omegan} denotes that the sequence, $h_n \defeq (2^n R)^{-3}\ln\Omega^{(n)}_{V^{(n)},N^{(n)}}(U^{(n)})$, is increasing about $n$, $h_{n-1} \le h_n$.
Therefore, $h_n$ diverges to $+\infty$ or converges to a {\em unique} constant, as $n \to \infty$.
On one hand, we find $h_n$ should be bounded above due to the proposition~\ref{prop:interacting}.
Hence, there exists a unique limit, $h_{\infty} = \lim_{n\to\infty}h_n$.

Note that, by definition of $V^{(n)}$, $N^{(n)}$ and $U^{(n)}$, we find $\lim\limits_{n\to\infty} N^{(n)}/V^{(n)} = N^{(0)}/R^3 \defeqr \rho$ and $\lim\limits_{n\to\infty} U^{(n)}/V^{(n)} = U^{(0)}/R^3 \defeqr \varepsilon$, where $\rho$ and $\varepsilon$ are constants.
This means that the limit operation, $n\to\infty$, of the large system considered here is the ``thermodynamic limit'' required in the statement of theorem.
Then, in this thermodynamic limit, the lower bound of $\varepsilon$ is given by, $\varepsilon \ge \varepsilon_g(\rho)$, due to $U^{(n)}\ge E^{(n)}_G(V^{(n)},N^{(n)})$ and the result~1 proven in step~2.
Furthermore, due to the limit $\lim\limits_{n\to\infty} (2^n R)^3/V^{(n)} = 1$, the existence of unique limit, $\lim\limits_{n\to\infty} (1/V^{(n)}\,)\ln\Omega^{(n)}_{V^{(n)},N^{(n)}}(U^{(n)}) = \lim\limits_{n\to\infty} h_n (2^n R)^3/V^{(n)}$, is obvious.
This is the thermodynamic limit, $\sigma(\varepsilon,\rho)$, given in Eq.\eref{eq:bf.thm.rt.2}.
Hence the aim of substep~3-1 is (roughly) achieved~\cite{note:proof.rt}.

%%%%%%%%%%
\subsubsection{Substep~3-2}

Consider the two systems in ${\mathcal D}^{(a)}$ and ${\mathcal D}^{(b)}$ introduced in step~1, which are not necessarily cubic.
Next, make $p_a$ copies of ${\mathcal D}^{(a)}$ including $N^{(a)}$ particles and $p_b$ copies of ${\mathcal D}^{(b)}$ including $N^{(b)}$ particles.
Consider the total system composed of these $p_a + p_b$ subsystems, and let the distance between arbitrary two subsystems is longer than or equal to $r_A$.
Then, the inequality~\eref{eq:bf.proof.rt.1.Omegatot} implies,
\eqb
 \Bigl[\, \Omega^{(a)}_{V^{(a)},N^{(a)}}\bigl(U^{(a)}\bigr) \,\Bigr]^{p_a} \cdot
 \Bigl[\, \Omega^{(b)}_{V^{(b)},N^{(b)}}\bigl(U^{(b)}\bigr) \,\Bigr]^{p_b}
 \le
 \Omega^{\rm tot}_{V^{\rm tot},N^{\rm tot}}\bigl( \widetilde{U} \bigr) \,,
\eqe
where $V^{\rm tot} = p_a V^{(a)} + p_b V^{(b)}$, $N^{\rm tot} = p_a N^{(a)} + p_b N^{(b)}$ and $\widetilde{U} = p_a U^{(a)} + p_b U^{(b)}$.
Take the logarithm and divide it by $V^{\rm tot}$,
\eqb
\label{eq:bf.proof.rt.3.Omegatot}
 \dfrac{\lambda}{V^{(a)}}\,
 \ln\Omega^{(a)}_{V^{(a)},N^{(a)}}\bigl(U^{(a)}\bigr)
 +
 \dfrac{1-\lambda}{V^{(b)}}\,
 \ln\Omega^{(b)}_{V^{(b)},N^{(b)}}\bigl(U^{(b)}\bigr)
 \le
 \dfrac{1}{V^{\rm tot}}\,
 \ln\Omega^{\rm tot}_{V^{\rm tot},N^{\rm tot}}\bigl( \widetilde{U} \bigr) \,,
\eqe
where $\lambda \defeq [\,1+(p_b V^{(b)}/p_a V^{(a)})\,]^{-1}$ which satisfies $0 < \lambda < 1$.
Here, consider the ``double'' thermodynamic limit, given by $V^{(a)}\to\infty$ with fixing $U^{(a)}/V^{(a)} = \varepsilon^{(a)}$ and $N^{(a)}/V^{(a)} = \rho^{(a)}$ at constant values, and $V^{(b)}\to\infty$ with fixing $U^{(b)}/V^{(b)} = \varepsilon^{(b)}$ and $N^{(b)}/V^{(b)} = \rho^{(b)}$ at constant values.
Then, we obtain from the inequality~\eref{eq:bf.proof.rt.3.Omegatot} and Eq.\eref{eq:bf.thm.rt.2} proven in substep~3-1,
\eqb
 \lambda\,\sigma(\varepsilon^{(a)},\rho^{(a)})
 + (1-\lambda)\,\sigma(\varepsilon^{(a)},\rho^{(a)})
 \le
 \sigma(\tilde{\varepsilon},\tilde{\rho}) \,,
\eqe
where $\tilde{\varepsilon} = \lambda\,\varepsilon^{(a)} + (1-\lambda)\,\varepsilon^{(b)}$ and $\tilde{\rho} = \lambda\,\rho^{(a)} + (1-\lambda)\,\rho^{(b)}$.
This is the same with Eq.\eref{eq:td.entropy.concave} and denotes that $\sigma(\varepsilon,\rho)$ is concave as a function of $\varepsilon$ and $\rho$.
The concavity is proven.

%%%%%%%%%%
\subsubsection{Substep~3-3}

It is obvious by definition of $\Omega_{V,N}(U)$ in Eq.\eref{eq:bf.Omega} that $\Omega_{V,N}(U)$ increases monotonously as $U$ increases.
Therefore, by definition of $\sigma(\varepsilon,\rho)$, it is obvious that $\sigma(\varepsilon,\rho)$ is monotone increasing about $\varepsilon$.
The (rough) proof of Ruelle-Tasaki theorem ends.
$\square$

%%%%%%%%%%%%%%%%%%%%%%%%%%%%%%%%%%%%%%%%%%%%%%%%%%%%%%%%%%%%%%%%%%%%%%%%%%%%%%%%%%%%%%%%%%%%%%%%%%%%
\section{Proof of Proposition~\ref{prop:interacting}}
\label{app:interacting}

The proof of proposition~\ref{prop:interacting} needs some preparations summarized in App.\ref{app:prep}.
As prepared in App.\ref{app:prep}, let $\epsilon_n$ be the energy eigen value of ``single-particle-state'' in an ideal gas which is a system with the interaction potential $\Phi = 0$ inside the system and $\Phi = \infty$ outside the system.
Then, by the condition~B of Ruelle-Tasaki theorem, it is easily found, $\bra{\psi} H_{V,N} \ket{\psi} \ge \bra{\psi} (\, H_{V,N}^{\rm (ideal)} - N \phi_B \,) \ket{\psi}$, where $H_{V,N}^{\rm (ideal)}$ is the Hamiltonian of ideal gas, and $\ket{\psi}$ is the arbitrary state in the intersection of Hilbert spaces, ${\mathcal H}_{V,N}\cap {\mathcal H}_{V,N}^{\rm (ideal)}$.
Then, the lemma~\ref{lemma:eigenvalue} given in App.\ref{app:prep} yields an inequality,
\eqb
\label{eq:interacting.1}
 E_k(V,N) \ge \epsilon_k - N\,\phi_B \quad, \text{for all $k$.}
\eqe
From this inequality, we can obtain a relation between the number of states of the interacting system, $\Omega_{V,N}(U)$, and that of the ideal gas, $\Omega_{V,N}^{\rm (ideal)}(U)$, as follows:

For a given integer $l$, the number of energy eigen values satisfying an inequality, $E_k(V,N) \le \epsilon_l-N\phi_B$, is expressed as $\Omega_{V,N}(\epsilon_l-N\phi_B)$ by definition.
This and the inequality~\eref{eq:interacting.1}, $E_k(V,N) \le \epsilon_l-N\phi_B \le E_l(V,N)$, denote that the number of states $\Omega_{V,N}(\epsilon_l-N\phi_B)$ is at most $l$.
On the other hand, using the number of states in ideal gas, we have $l = \Omega_{V,N}^{\rm (ideal)}(\epsilon_l)$.
Hence, we find $\Omega_{V,N}(\epsilon_l-N\phi_B) \le \Omega_{V,N}^{\rm (ideal)}(\epsilon_l)$.
Then, by introducing $U$ as $U \defeq \epsilon_l-N\phi_B$ and using the lemma~\ref{lemma:ideal} given in App.\ref{app:prep}, we obtain,
\eqb
 \Omega_{V,N}(U) \le
 \Omega_{V,N}^{\rm (ideal)}(U+N\phi_B) \le
 \exp\bigl[\, \tilde{\sigma}_0\,V + \tilde{\beta}\,(U+N\phi_B) \,\bigr] \,,
\eqe
where $\tilde{\sigma}_0$ is a constant introduced in lemma~\ref{lemma:ideal}.
By introducing a constant, $\tilde{\sigma} = \tilde{\sigma}_0 + \tilde{\beta} \rho \phi_B$, the proposition~\ref{prop:interacting} is proven.
$\square$

%%%%%%%%%%%%%%%%%%%%%%%%%%%%%%%%%%%%%%%%%%%%%%%%%%%%%%%%%%%%%%%%%%%%%%%%%%%%%%%%%%%%%%%%%%%%%%%%%%%%
\section{Preparations for Proposition~\ref{prop:interacting}}
\label{app:prep}

This appendix shows two lemmas as the preparation of the proof of proposition~\ref{prop:interacting}.
The first lemma is a consequence of the mini-max principle (proposition~\ref{prop:minimax}):
\begin{lemma}
\label{lemma:eigenvalue}
Suppose that there are two Hamiltonians, $H_{V,N}^{(1)}$ and $H_{V,N}^{(2)}$, which differ by the interaction potential but the system volume and particle number are the same.
Let, $E_k^{(i)}(V,N)$ ($i=1,\,2$ and $k=1,\,2\,\cdots$) be the energy eigen value of each Hamiltonian, and $k$ is attached in increasing order $E_k^{(i)} \le E_{k+1}^{(i)}$.
Under this presupposition, if the inequality, $\bra{\psi} H_{V,N}^{(1)} \ket{\psi} \le \bra{\psi} H_{V,N}^{(2)} \ket{\psi}$, holds for all states $\ket{\psi}$ in the intersection of Hilbert spaces (\,$^\forall\ket{\psi} \in {\mathcal H}_{V,N}^{(1)}\cap{\mathcal H}_{V,N}^{(2)}$\,), then the inequality of eigen value, $E_k^{(1)}(V,N) \le E_k^{(2)}(V,N)$, holds for all $k$.
\end{lemma}
Proof of this lemma is found in textbooks of functional analysis and mathematical foundation of quantum mechanics.

Before showing the next lemma, let us summarize ground partition function and ground potential.
For the quantum system considered in Ruelle-Tasaki theorem, the ground partition function, $\Xi_{V,\beta,\mu}$, is defined by
\eqb
\label{eq:prep.Xi}
 \Xi_{V,\beta,\mu} \defeq
 \sum_{N=0}^{\infty} \exp\bigl(\beta \mu N\bigr)\,
 {\rm Tr}\, \exp\bigl( -\beta H_{V,N} \bigr)
 =
 \sum_{N=0}^{\infty}\,\exp\bigl( \beta \mu N \bigr)\,
 \sum_{k=1}^{\infty}\,\exp\bigl[ -\beta E_k(V,N) \bigr] \,,
\eqe
where $\beta \defeq T^{-1}$ is the inverse of temperature, and $\mu$ is the chemical potential.
The {\em density of ground potential} at large system limit, $q_{\beta,\mu}$, is defined by $q_{\beta,\mu}(\rho) \defeq \lim_{l.s.l}[\, - (1/\beta V)\ln\Xi_{V,\beta,\mu}\,]$, where $\lim_{l.s.l.}$ means the large system limit defined in the statement of Ruelle-Tasaki theorem, and $\rho = N/V$ is the number density fixed at constant in the limit operation.
In ordinary thermodynamics, $q_{\beta,\mu}$ corresponds to the minus of pressure.

If the system is an ideal gas (i.e. $\Phi = 0$ inside the system, and $\Phi = \infty$ outside the system), then the ground partition function becomes,
\eqb
\label{eq:prep.Xi.ideal}
 \Xi_{V,\beta,\mu}^{\rm (ideal)} = \prod_{n=1}^{\infty}\,\xi_{\beta,\mu}(\epsilon_n) \quad,\quad
 \xi_{\beta,\mu}(\epsilon_n) =
 \begin{cases}
  1 + e^{-\beta(\epsilon_n - \mu)} & \text{:\,\,fermionic ideal gas} \\
  [\,1 - e^{-\beta(\epsilon_n - \mu)}\,]^{-1} & \text{:\,\,bosonic ideal gas}
 \end{cases} \,,
\eqe
where $\epsilon_n$ is the energy eigen value of ``single-particle-state'' of ideal gas.
The density of ground potential at large system limit becomes,
\eqb
\label{eq:prep.q.ideal}
 q^{\rm (ideal)}_{\beta,\mu}(\rho)
 = - \lim_{l.s.l.} \dfrac{1}{\beta V}\,\ln\Xi^{\rm (ideal)}_{V,\beta,\mu}
 = - \dfrac{1}{\beta}
   \int_{\epsilon_g}^{\infty} d\epsilon\,\nu(\epsilon)\,\ln\xi_{\beta,\mu}(\epsilon) \,,
\eqe
where $\epsilon_g$ is the ground state energy of the ideal gas, $\mu < \epsilon_g$ is assumed for bosonic gas, and $\nu(\epsilon)$ is the number of single-particle-states per energy interval $d\epsilon$ per unit volume.
($\nu(\epsilon) \propto \epsilon^{1/2}$ for spatially three dimensional case.)
The important fact in this appendix is that the integral in Eq.\eref{eq:prep.q.ideal} converges for both fermionic and bosonic ideal gases.
Using this fact, let us show the following lemma:
\begin{lemma}
\label{lemma:ideal}
Let $\Omega^{(ideal)}_{V,N}(U)$ be the number of states~\eref{eq:bf.Omega} for an ideal gas .
Then, for arbitrary constants, $\tilde{\beta}\,(>0)$ and $\tilde{\mu}\,(<\epsilon_g\,\,\text{for bosonic gas})$, the number of states $\Omega_{V,N}^{\rm (ideal)}(U)$ is bounded above at the large system limit,
\eqb
\label{eq:prep.ideal.1}
 \Omega^{\rm (ideal)}_{V,N}(U) \le
 \exp\bigl(\, \tilde{\sigma}_0\,V + \tilde{\beta}\,U \,\bigr) \,,
\eqe
where $\tilde{\sigma}_0(\rho,\tilde{\beta},\tilde{\mu})$ is a constant, and $\rho = N/V$ is the number density fixed in the large system limit.
\end{lemma}

\noindent
[{\em Proof of lemma~\ref{lemma:ideal}}] 
By definition~\eref{eq:bf.Omega}, $\Omega_{V,N}^{\rm (ideal)}(U)$ is expressed as
\eqb
 \Omega_{V,N}^{\rm (ideal)}(U) =
 \sum_{\gamma}\,\,
 \chi\Bigl[\,\sum_{n=1}^{\infty}n_n = N\,\Bigr] \,\,
 \chi\Bigl[\,\sum_{n=1}^{\infty}\epsilon_n n_n \le U\,\Bigr] \,,
\eqe
where $\chi[ \text{equation} ] = 1$ if ``equation'' holds and $\chi[ \text{equation} ] = 0$ if ``equation'' does not hold, $n_n$ is the number of particles at $n$-th energy level of single-particle-state, and $\gamma = \{\, n_1 \,,\, n_2 \,\cdots \,\}$ is the distribution of particles in all energy levels.
By the explicit relations, $e^x > 0$ for all $x$ and $e^x \ge 1$ for $x \ge 0$, we find,
\eqb
\begin{array}{rcl}
 \Omega_{V,N}^{\rm (ideal)}(U)
 &\le
 &\displaystyle
  \sum_{\gamma}\,
  \exp\Bigl[\,
  - \tilde{\beta}\,\Bigl\{\,\sum_{n=1}^{\infty}\epsilon_n n_n - U \,\Bigr\}
  + \tilde{\beta}\,\tilde{\mu}\,\Bigl\{\,\sum_{n=1}^{\infty}n_n - N \,\Bigr\}
  \,\Bigr] \\
 &=
 &\displaystyle
  e^{\tilde{\beta}\,(U-\tilde{\mu}N)}\,
  \sum\limits_{\gamma}\,
  \exp\Bigl[\,
  - \tilde{\beta}\,\Bigl\{\,\sum_{n=1}^{\infty}\epsilon_n n_n\,\Bigr\}
  + \tilde{\beta}\,\tilde{\mu}\,\Bigl\{\,\sum_{n=1}^{\infty}n_n \,\Bigr\}
  \,\Bigr] \,.
\end{array}
\eqe
Then, by the standard calculation of ground potential in statistical mechanics, we obtain,
\eqb
 \Omega_{V,N}^{\rm (ideal)}(U) \le
 e^{\tilde{\beta}\,(U-\tilde{\mu}N)}\,\Xi_{V,\tilde{\beta},\tilde{\mu}}^{\rm (ideal)} \,,
\eqe
where $\Xi_{V,\tilde{\beta},\tilde{\mu}}^{\rm (ideal)}$ is given in Eq.\eref{eq:prep.Xi.ideal}.
This inequality together with Eq.\eref{eq:prep.q.ideal} yield the following inequality at the large system limit,
\eqb
 \Omega_{V,N}^{\rm (ideal)}(U) \le
 \exp\bigl[\, \tilde{\beta}\,( U - \tilde{\mu} N )
            - V \tilde{\beta}\, q_{\tilde{\beta},\tilde{\mu}}^{\rm (ideal)}(\rho)
 \,\bigr] \,,
\eqe
where $\tilde{\mu} < \epsilon_g$ is required for bosonic ideal gas as mentioned at Eq.\eref{eq:prep.q.ideal}.
Hence, by introducing a constant, $\tilde{\sigma}_0 = -\tilde{\beta} \tilde{\mu} \rho - \tilde{\beta} q_{\tilde{\beta},\tilde{\mu}}^{\rm (ideal)}(\rho)$, the lemma~\ref{lemma:ideal} is proven.
$\square$

%%%%%%%%%%%%%%%%%%%%%%%%%%%%%%%%%%%%%%%%%%%%%%%%%%%%%%%%%%%%%%%%%%%%%%%%%%%%%%%%%%%%%%%%%%%%%%%%%%%%
\section{Proof of Corollary~\ref{coro:bf}}
\label{app:coro}

Given the Ruelle-Tasaki theorem, we find for a sufficiently large $V$,
\eqb
\begin{array}{rcl}
 \dfrac{\Omega_{V,N}(U-\delta V)}{\Omega_{V,N}(U+\delta V)}
 & =
 & \exp\Bigl[\,
        V\,\bigl\{\, \sigma(\varepsilon-\delta,\rho)
              - \sigma(\varepsilon+\delta,\rho) \,\bigr\} + O(V^q) \,\Bigr] \quad, (\,q<1\,)\\
 & =
 & \exp\Bigl[\,
        V\,\Bigl\{\, -\pd{\sigma(\varepsilon,\rho)}{\varepsilon}\,2\,\delta
                 + O(\delta^3) \,\Bigr\} + O(V^q) \,\Bigr]
  \,\, \to \,\,0 \quad \text{as $V\to\infty$}\,.
\end{array}
\eqe
Therefore, by definition of $W_{V,N}(U,\delta)$, we find
\eqb
\begin{array}{rcl}
 W_{V,N}(U,\delta)
 & =
 & \Omega_{V,N}(U+\delta V) - \Omega_{V,N}(U-\delta V) \\
 & =
 & \Omega_{V,N}(U+\delta V)\,
   \Bigl[\,1 - \dfrac{\Omega_{V,N}(U-\delta V)}{\Omega_{V,N}(U+\delta V)} \,\Bigr]
  \,\, \to \,\,\Omega_{V,N}(U+\delta V) \quad \text{as $V\to\infty$}\,.
\end{array}
\eqe
Hence, replacing $U+\delta V$ with $U$, we obtain Eq.\eref{eq:bf.bf} from Eq.\eref{eq:bf.bf.usual} at thermodynamic limit.
$\square$

%%%%%%%%%%%%%%%%%%%%%%%%%%%%%%%%%%%%%%%%%%%%%%%%%%%%%%%%%%%%%%%%%%%%%%%%%%%%%%%%%%%%%%%%%%%%%%%%%%%%
%==========================================================
% Back Matter (References and Notes)
%----------------------------------------------------------
% Style and layout of the references

\bibliographystyle{mdpi}
\makeatletter
\renewcommand\@biblabel[1]{#1. }
\makeatother

%----------------------------------------------------------
% Use the following option to include external BibTeX files:
% \bibliography{template}

\begin{thebibliography}{99}
%
\bibitem{ref:bardeen+2.1973}
 Bardeen, J.M.; Carter, B.; Hawking, S.W.
 The Four Laws of Black Hole Mechanics.
 {\em Commun.Math.Phys.} {\bf 1973}, {\em 31}, 161-170.
%
\bibitem{ref:bekenstein.1973}
 Bekenstein, J.D.
 Black Holes and Entropy.
 {\em Phys.Rev.} {\bf 1973}, {\em D7}, 2333-2346.
%
\bibitem{ref:bekenstein.1974}
 Bekenstein, J.D.
 Generalized second law of thermodynamics in black-hole physics.
 {\em Phys.Rev.} {\bf 1974}, {\em D9}, 3292-3300.
%
\bibitem{ref:braden+3.1990}
 Braden, H.W.; Brown, J.D.; Whiting, B.F.; York, J.W.,Jr.
 Charged black hole in a grand canonical ensemble.
 {\em Phys.Rev.} {\bf 1990}, {\em D42}, 3376-3385.
%
\bibitem{ref:brown+2.1991}
 Brown, J.D.; Martinez, E.A.; York, J.W.,Jr.
 Complex Kerr-Newman Geometry and Black-Hole Thermodynamics.
 {\em Phys.Rev.Lett.} {\bf 1991}, {\em 66}, 2281-2284.
%
\bibitem{ref:davies.1977}
 Davies, P.C.W.
 The thermodynamics theory of black holes.
 {\em Proc.R.Soc.Lond.} {\bf 1977}, {\em A353}, 499-521.
%
\bibitem{ref:hawking.1971}
 Hawking, S.W.
 Gravitational Radiation from Colliding Black Holes.
 {\em Phys.Rev.Lett.} {\bf 1971}, {\em 26}, 1344-1346.
%
\bibitem{ref:hawking.1975}
 Hawking, S.W.
 Particle Creation by Black Holes.
 {\em Commun.Math.Phys.} {\bf 1975}, {\em 43}, 199-220.
%
\bibitem{ref:israel.1986}
 Israel, W.
 Third Law of Black-Hole Dynamics: A Formulation and Proof.
 {\em Phys.Rev.Lett.} {\bf 1986}, {\em 57}, 397-399.
%
\bibitem{ref:york.1986}
 York, J.W.,Jr.
 Black-hole thermodynamics and the Euclidean Einstein action.
 {\em Phys.Rev.} {\bf 1986}, {\em D33}, 2092-2099.
%
\bibitem{ref:flanagan+2.2000}
 Flanagan, E.E.; Marolf, D.; Wald, R.M.
 Proof of Classical Version of the Bousso Entropy Bound and of the Generalized Second Law.
 {\em Phys.Rev.} {\bf 2000}, {\em D62}, 084035:1-12.
%
\bibitem{ref:frolov+1.1993}
 Frolov, V.P. ; Page, D.N.
 Proof of the Generalized Second Law for Quasistationary Semiclassical Black Holes.
 {\em Phys.Rev.Lett.} {\bf 1993}, {\em 71}, 3902-3905.
%
\bibitem{ref:saida.2006}
 Saida, H.
 The generalized second law and the black hole evaporation in an empty space as a nonequilibrium process.
 {\em Class.Quant.Grav.} {\bf 2006}, {\em 23}, 6227-6243.
%
\bibitem{ref:unruh+1.1982}
 Unruh, W.G.; Wald, R.M.
 Accelerated radiation and the generalized second law of thermodynamics.
 {\em Phys.Rev.} {\bf 1982}, {\em D25}, 942-958.
 (Correction in {\em Phys.Rev.} {\bf 1988}, {\em D37}, 3059-3060.)
%
\bibitem{ref:unruh+1.1983}
 Unruh, W.G.; Wald, R.M.
 Entropy bounds, acceleration radiation, and the generalized second law.
 {\em Phys.Rev.} {\bf 1983}, {\em D27}, 2271-2276.
%
\bibitem{ref:saida.2009b}
 Saida, H.
 To what extent is the entropy-area law universal? -- Multi-horizon and multi-temperature spacetime may break the entropy-area law --.
 {\em Prog.Theor.Phys.} {\bf 2009}, {\em 122}, 1515-1552.
%
\bibitem{ref:birrell+1.1982}
 Birrell, N.D.; Davies, P.C.W.
 {\em Quantum fields in curved space};
 Cambridge Univ. Press: Cambridge, UK, 1982.
%
\bibitem{ref:bousso.2002}
 Bousso, R.
 The holographic principle.
 {\em Rev.Mod.Phys.} {\bf 2002}, {\em 74}, 825-874.
%
\bibitem{ref:lieb+1.1999}
 Lieb, E.H.; Yngvason, J.
 The Physics and Mathematics of the Second Law of Thermodynamics.
 {\em Phys.Rep.} {\bf 1999}, {\em 310}, 1-96.
%
\bibitem{ref:tasaki.2000}
 Tasaki, H.
 {\em Thermodynamics (Netsu-Rikigaku)};
 Baifu-Kan Publ.: Tokyo, Japan, 2000 (Published in Japanese).
%
\bibitem{ref:dobrushin.1964}
 Dobrushin, R.L.
 Investigation of the Conditions of the Asymptotic Existence of the Configuration Integral of the Gibbs Distribution.
 {\em Teorija Verojatn. i ee Prim.} {\bf 1964}, {\em 9}, 626-643.
%
\bibitem{ref:ruelle.1999}
 Ruelle, D.
 {\em Statistical Mechanics: Rigorous Results};
 Imperial College Press and World Scientific Publ.: London and Singapore, UK and Singapore, 1999;
 Chapter 1-3.
%
\bibitem{ref:tasaki.2008}
 Tasaki, H.
 {\em Statistical Mechanics (Tokei-Rikigaku)};
 Baifu-Kan Publ.: Tokyo, Japan, 2008 (Published in Japanese).
%
\bibitem{note:principle.1}
The increase of temperature by this adiabatic process is regarded as one basic principle in the axiomatic thermodynamics.
%
\bibitem{note:entropy.principle}
For example, the Kelvin's statement of second law, the basic principles mentioned in notes~\cite{note:principle.1} and~\cite{note:principle.2}, and so on.
%
\bibitem{note:principle.2}
One of basic principles of axiomatic thermodynamics is the existence of an adiabatic process which connects arbitrary two thermal equilibrium states.
The ``direction'' of adiabatic process is to be determined by the entropy principle.
%
\bibitem{note:temp/mass}
The ratio of Hawking temperature, $\hbar c^3/8\pi G M$ given in Eq.\eref{eq:td.Tbh} at spatial infinity ($r_w \to \infty$), to black hole mass energy, $M c^2$, is $\hbar c/8\pi G M_{\odot}^2 \simeq (\sqrt{2}\,{\rm g})^2 M^{-2}\times 10^{-11}$.
This ratio is of order unity for Planck mass, $M \simeq 2\times 10^{-5}$~g.
For solar mass black hole, $M_{\odot} \simeq 2\times 10^{33}$~g, this ratio is of order $10^{-77}$.
%
\bibitem{note:equilibrium}
In general, global thermal equilibrium state does not evolve in time by its definition.
In black hole thermodynamics, we consider the global equilibrium of the system shown in Fig.\ref{fig:2}, and thus the static coordinate is suitable.
Here, we should distinguish the ``global'' equilibrium and ``local'' equilibrium.
For example, a fluid can be in local equilibrium, in which each fluid element is in a thermal equilibrium state but the equilibrium state of one fluid element is not necessarily the same with that of the other element.
The fluid is globally in non-equilibrium states, since the states of fluid elements do not necessarily balance with each other.
The fluid evolves in time due to this global non-equilibrium nature.
%
\bibitem{ref:tolman.1987}
 Tolman, R.C.
 {\em Relativity, Thermodynamics and Cosmology};
 Dover Publ.: New York, USA, 1987; Chapter 9.
%
\bibitem{ref:gibbons+1.1977}
 Gibbons, G.W.; Hawking, S.W.
 Action integrals and partition functions in quantum gravity.
 {\em Phys.Rev.} {\bf 1977}, {\em D15}, 2752-2256.
%
\bibitem{ref:hawking.1979}
 Hawking, S.W.
 The Path-Integral Approach to Quantum Gravity.
 In {\em Euclidean Quantum Gravity}; Gibbons, G.W.; Hawking, S.W., Eds.; World Scientific Publ.:
 Singapore, 1993.
%
\bibitem{ref:saida.2009a}
 Saida, H.
 de~Sitter thermodynamics in the canonical ensemble.
 {\em Prog.Theor.Phys.} {\bf 2009}, {\em 122} 1239-1266.
%
\bibitem{ref:fulling+1.1976}
 Fulling, S.A.; Davies, P.C.W.
 Radiation from a moving mirror in two dimensional space-time: conformal anomaly.
 {\em Proc.R.Soc.Lond.} {\bf 1976}, {\em A348}, 393-414.
%
\bibitem{ref:martinez+1.1989}
 Martinez, E.A.; York, J.W.,Jr.
 Additivity of the entropies of black holes and matter in equilibrium.
 {\em Phys.Rev.} {\bf 1989}, {\em D40}, 2124-2127.
%
\bibitem{note:realsystem}
One may consider, for example, the electron gas as an example of the system which violates the condition~A but retains the Boltzmann formula, since the interaction potential between electrons is positive at large distances and violates the condition~A. 
However let us note that, in Ruelle's book~\cite{ref:ruelle.1999}, the condition~A is extended so as to include the interaction potential which can become positive and repulsive at large distances. 
(Note that it is sufficient for the aim of this paper to consider the case that the potential is negative at large distances, such as the Newtonian gravity.) 
Hence, by such an extension of condition~A, the electron gas can be regarded as the system which satisfies the sufficient condition for the results~1 and~2 in Ruelle-Tasaki theorem.
%
\bibitem{note:dobrushin}
The author could not obtain the original paper~\cite{ref:dobrushin.1964} of Dobrushin.
But this theorem is found in Ruelle's book~\cite{ref:ruelle.1999} as proposition 3.2.4.
The Dobrushin theorem in Ruelle's book is proven only for classical systems.
However, we are interested in quantum system in this paper.
Therefore, the statement and proof of Dobrushin theorem in this paper are the extended version by this author so as to match with quantum system under consideration.
%
\bibitem{ref:iyer+1.1994}
 Iyer, V.; Wald, R.M.
 Some properties of the Noether charge and a proposal for dynamical black hole entropy.
 {\em Phys.Rev.} {\bf 1994}, {\em D50}, 846-864
%
\bibitem{note:proof.rt}
We have used a special large system composed of the cubes ${\mathcal D}^{(n)}$.
The extension of this special system to the general large system is found in Ruelle's book~\cite{ref:ruelle.1999}.
The other mathematical details, such as the uniformity of convergence of $\varepsilon_g(\rho)$ about $\rho$, are also found in Ruelle's book.
%
\end{thebibliography}
%----------------------------------------------------------

%%%%%%%%%%%%%%%%%%%%%%%%%%%%%%%%%%%%%%%%%%%%%%%%%%%%%%%%%%%%%%%%%%%%%%%%%%%%%%%%%%%%%%%%%%%%%%%%%%%%
\end{document}